\documentclass[aps,reprint,prb,groupedaddress,showpacs,superscriptaddress]{revtex4-1}
\usepackage{amsmath,amssymb,amsthm,color}
\usepackage{graphicx}
\usepackage{ulem}
\usepackage{hyperref}

\begin{document}

\title{Josephson junction of finite-size superconductors on a topological insulator under a magnetic field}

%\title{Finite-size superconductor-topological insulator-superconductor Josephson junction}

%{Dynamics of Josephson effects in a Planar Superconductor-Topological Insulator-Superconductor Heterostructure isolated by topologically trivial gap.}

\author{Sang-Jun Choi}
\affiliation{Department of Physics, Korea Advanced Institute of Science and Technology, Daejeon 34141, Korea}
\affiliation{\mbox{Center for Theoretical Physics of Complex Systems, Institute for Basic Science, Daejeon 34126, Korea}}
\author{H.-S. Sim} \email[Corresponding author.]{hssim@kaist.ac.kr}
\affiliation{Department of Physics, Korea Advanced Institute of Science and Technology, Daejeon 34141, Korea}
\date{\today}

\begin{abstract} 
We theoretically study a Josephson junction formed by two finite-size $s$-wave SCs on a topological insulator under a magnetic field.
At certain conditions, the junction hosts the chiral Majorana modes enclosing the two finite-size SCs. The interplay of the extended chiral Majorana modes and the states inside the junction can results in nontrivial topological effects such as  the $2n \pi$ fractional AC Josephson effects predicted in Ref.~\cite{ChoiSim}
%We find the condition for the emergence of the extended chiral Majorana zero modes along the boundary of the SC regions outside the junction. 
We show that the $2n \pi$ fractional AC Josephson effects can occur in a realistic situation, such as the presence of the midgap states, without requiring fine tuning of the parameters of the junction.
%Moreover, we show that the $2n \pi$ fractional AC Josephson effect is insensitive to continuous deformation of the setup such as the shape of the arcs and different arc-lengths of the SCs. 
%Although the $2n\pi$ fractional AC Josephson effect offers the simplest way of observing the non-Abelian evolution of the Josephson junction, we provide another approach which offers a more direct way of observing the non-Abelian nature of the Majorana zero mode.
We also find that the Shapiro spikes of the junction show a rich structure  in a wide range of the AC voltage bias, facilitating experimental identification of the $2n \pi$ fractional AC Josephson effects.  
Moreover, we discuss how to observe the non-commutativity of the operations that braid the Majorana fermions of the junction, by measuring the Josephson current.
Finally, we study the state evolution of the junction when the junction hosts a different number of Majorana zero modes from the case of the $2n \pi$ fractional AC Josephson effects. 
\end{abstract}

\maketitle

\section{Introduction}
Implementation of topological quantum computation requires exotic quasi-particles obeying non-Abelian braiding statistics.~\cite{Nayak,Stern}
Majorana fermions are non-Abelian anyons,~\cite{Wilczek,Alicea,Flensberg} and experimental observation of their non-Abelian effects would be an important step towards topological quantum computation.
Majorana fermions emerge as topological zero-energy excitations of $p$-wave superconductors (SCs) in which they come in pair as the real and imaginary parts of a complex fermion.
There are the effective $p$-wave superconductors that can host Majorana fermions,~\cite{Lutchyn, Oreg, FuKane08, Yazdani, Qi, ShtengelPRX,Kouwenhoven, Heiblum,Deng,Hao} including a Josephson junction of $s$-wave superconductors on a topolgical insulator (TI).~\cite{PotterFu,twoMFs,Jin,Pientka,Sun}
%A huge amount of researches toward harnessing Majorana fermions has been sparked in a variety of heterostructures imitating  effectively the $p$-wave SCs

%As an enthusiasm identifying Majorana fermions, $4\pi$ fractional AC Josephson effect has been pursued with the Josephson junctions on one-dimensional systems such as edge channels of quantum spin Hall insulators~\cite{Wiedenmann,Furdyna}. 
%As an enthusiasm identifying Majorana fermions, $4\pi$ fractional AC Josephson effect has been pursued with the Josephson junctions on one-dimensional conducting channels~\cite{Wiedenmann,Furdyna,Molenkamp,Laroche}. 
%Conventional Josephson junctions having $2\pi$-period in terms of superconducting phase difference show the oscillating  supercurrent with the Josephson frequency $f_\text{J}$ under voltage bias. 

The $4 \pi$ fractional AC Josephson effect is one of the key features of the $p$-wave topological SC.~\cite{Kitaev1D,FuKanePRB}
Conventional Josephson junctions transfer a pair of electron charge when the SC phase difference changes by $2\pi$, showing the supercurrent oscillating in time with the Josephson frequency $\omega_\text{J}=2eV_\text{DC}/\hbar$ under voltage bias $V_\text{DC}$ across the junction.
By contrast, in topological Josephson junctions hosting two Majorana fermions,
a single electron can tunnel into a junction state (e.g., $|0\rangle$, formed by the two Majorana fermions) and change the occupation-number parity of the state
(e.g., forming $|1 \rangle$), when the phase difference changes by $2\pi$. Due to this fermion parity switching, the initial junction state is recovered  after the SC phase difference changes by $4\pi$. This results in the fractional Josephson frequency of $\omega_\text{J}/2$, namely, the period elongation by factor two.

The Shapiro step or spike is a useful tool for observing the frequency of oscillating supercurrent from long-time average of the voltage or current across the junction.
By measuring the Shapiro effect, the $4 \pi$ fractional AC Josephson effect has been observed.~\cite{Wiedenmann,Furdyna,Molenkamp,Laroche}
In a recent experiment~\cite{Molenkamp}, the Shaprio spikes were detected in a Josephson junction on a TI by applying DC bias voltage to the junction and measuring the Josephson radiation, and the fractional frequency $\omega_\text{J}/2$ was identified.
 
%, as it requires only average value of the oscillating current, i.e., DC measurement, for a much longer time. The Josephson radiation experiment has measured the fractional frequency $\omega_\text{J}/2$ of oscillating supercurrent under $V_\text{DC}$ demonstrating the $4\pi$ fractional Josephson effect~\cite{Molenkamp} with HgTe. In another recent experiment~\cite{Laroche} in which a topological Josephson junction is exposed to monochromatic radiation with a frequency $\omega_0$, sharp increase of DC supercurrent is observed when $\omega_0$ matches to the fractional frequency of $\omega_\text{J}/2$. 

%Recently, Ref.~\cite{ChoiSim} has proposed a single Josephson junction hosting four Majorana fermions under a magnetic field, enabling the $2\pi$ fractional Josephson effect due to the braiding of Majorana fermions in a setup which can come into realization in near future.  

Recently,  the $2n\pi$ fractional Josephson effect with an integer $n \ge 2$ was theoretically predicted in Ref.~\cite{ChoiSim}.
The effect shows the Josephson current oscillating with the fractional frequency of $\omega_\textrm{J} / n$ in time when a DC bias voltage $V_\text{DC}$ is applied across the junction. The period elongation factor $n$ is an integer $\ge 2$ and 
can be tuned by the DC bias voltage. This is a unique feature of the $2n \pi$ fractional AC Josephson effect. The $2n \pi$ fractional effect occurs in a Josephson junction formed by two finite-size $s$-wave SCs on a TI when three magnetic flux quanta are applied through the junction  (see Fig.~\ref{Fig:setup}).
The $2n \pi$ fractional Josephson effect can be observed by measuring Shapiro spikes.

The origin of the $2n\pi$ fractional Josephson effect is the fusion and braiding of four Majorana fermions formed in the junction.
The fusion and braiding results in non-Abelian state evolution such as 
$| 00 \rangle \to |00\rangle+e^{i\phi}|11\rangle$ in one conventional Josephson period $T_\textrm{J} = h / (2e V_\textrm{DC}) = 2\pi / \omega_\text{J}$, where $|00 \rangle$ and $|11 \rangle$ are the complex fermion states formed by the four Majorana fermions. The evolution turns back to the initial state after time $n T_\textrm{J}$.
The dynamical phase $\phi$ is accumulated during the fusion and depends on  the voltage $V_\textrm{DC}$, and this is why the period-elongation factor $n$ or the state evolution can be tuned by changing $V_\textrm{DC}$.
This mechanism is different from that of the known $4 \pi$ fractional Josephson effect~\cite{Kitaev1D,FuKanePRB} where only two Majorana fermions play a role and no non-Abelian state evolution takes place.

One interesting point is that the non-Abelian state evolution occurs in the single Josephson junction; this contrasts with the other theoretical proposals~\cite{FuKane08,SternBerg} that require multiple junctions to generate non-Abelian state evolution by braiding Majorana fermions. One of the key elements that make the non-Abelian evolution possible in the single junction is the
geometrical feature that the two SC regions of the junction are of finite size.
Due to a topological orgin, the junction hosts the chiral Majorana modes enclosing the two finite-size SC regions. The interplay of the extended chiral Majorana modes and the states inside the junction can generate new effects that are absent in conventional Josephson junctions. This motivates us to study the various properties of the junction in this work.

In this paper, we develop the theory for the Josephson junction formed by two finite-size $s$-wave SCs on a topological insulator under a magnetic field  (see Fig.~\ref{Fig:setup}).
We derive the effective Hamiltonian describing the low-energy phenomena of the junction, starting from the Dirac-Bogoliubov-de Gennes Hamiltonian, and find the condition for the emergence of the extended chiral Majorana zero modes along the boundary arcs of the SC regions outside the junction. 
Extending the findings in Ref.~\cite{ChoiSim}, we show that the $2n \pi$ fractional AC effects do not require fine tuning of the parameters and that the $2n \pi$ effects can occur in the setup with realistic parameters in the presence of the midgap states of the junction. 
%Moreover, we show that the $2n \pi$ fractional AC Josephson effect is insensitive to continuous deformation of the setup such as the shape of the arcs and different arc-lengths of the SCs. 
Moreover, we discuss how to observe the non-commutativity of the operations that braid the Majorana fermions of the junction.
%Although the $2n\pi$ fractional AC Josephson effect offers the simplest way of observing the non-Abelian evolution of the Josephson junction, we provide another approach which offers a more direct way of observing the non-Abelian nature of the Majorana zero mode.
We also compute the Shapiro spikes of the $2n \pi$ fractional AC Josephson effect  in a wide range of a AC voltage bias across the junction and show that the spikes have a richer structure than those at a small AC voltage studied in Ref.~\cite{ChoiSim}, facilitating more possibility of experimental identification of the $2n \pi$ fractional AC Josephson effects.  
Finally, we study how the state evolution of the junction or the $2n \pi$ fractional AC Josephson effect changes when the junction hosts a different number of Majorana zero modes inside the junction or inside the SC regions than the case studied in Ref.~\cite{ChoiSim}.

% the various number of Majorana fermions with the different number of magnetic flux quanta inside the junction. We also study the effects of the Abirokosov vortices on the $2 n \pi$ fractional AC Josephson effect in the case of $N=2,3$. We find that a modified $2n \pi$ fractional Josephson effect occurs with any number of Abrikosov vortices, if a suitable strength of the magnetic field is chosen.

This paper is organized as follows. 
In Sec.~\ref{Sec:Model}, we introduce the Josephson junction. 
We derive the low-energy description of the junction, study the emergence of Majorana fermions, and briefly review the $2n\pi$ fractional Josephson effect found in Ref.~\cite{ChoiSim}. 
In Sec.~\ref{Sec:Practical}, we discuss the robustness of the $2n\pi$ fractional Josephson effect, and provide the numerical results of the Josephson current with realistic parameters in the presence of the midgap states.
In Sec.~\ref{Sec:NonCom}, we study how to observe the non-commutativity of the braiding operations.
In Sec.~\ref{Sec:Shapiro}, we numerically compute the Shapiro spikes.
In Sec.~\ref{Sec:MultipleMFs}, we discuss the cases where the junction hosts  a different number of Majorana zero modes from the case studied in Ref.~\cite{ChoiSim}.
The summary is given in Sec.~\ref{Sec:Summary}.

%Although the recent experiments on the Shapiro steps successfully have demonstrated the $4\pi$ fractional Josephson effects, it has been followed by further theoretical studies that topologically trivial Josephson junctions without Majorana fermions are able to exhibit the fractional Josephson effect caused by the fermion parity switching~\cite{Egger,DasSarma,Mora}. These render the demonstration of Majorana fermions elusive.

\section{Setup}\label{Sec:Model}
\begin{figure}[t]
\centering
\includegraphics[scale=0.55]{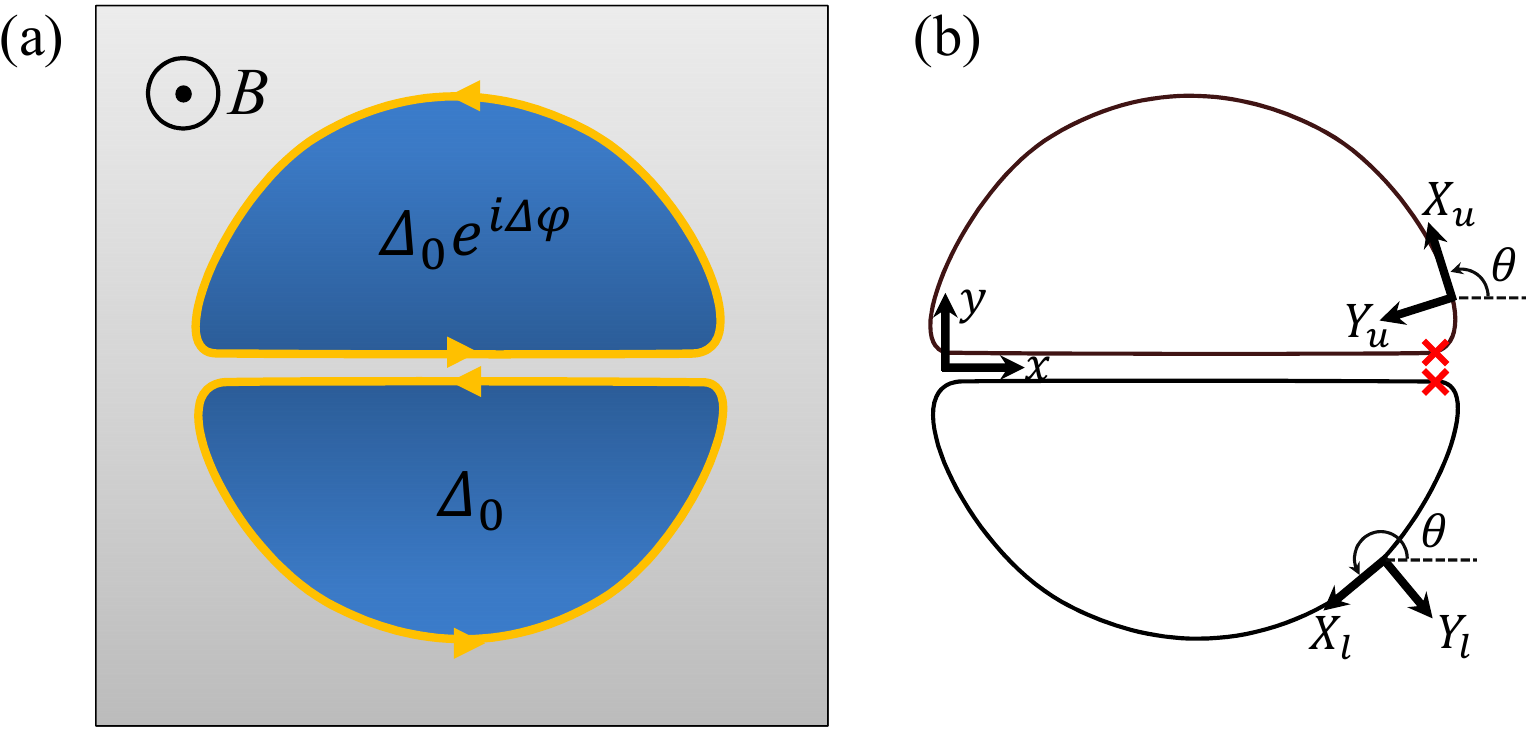}
\caption{ (a)  A Josephson junction consists of two finite-size effective $p$-wave superconducting regions (blue shade) on a TI.  A magnetic field $B$ is applied to the TI, to make the surface region outside the SCs insulating.
%egion outside the SCs topologically trivial insulating region (grey shade) under a magnetic field $B$.
Chiral Majorana modes (yellow lines) propagate along the boundary of each SC region in the direction represented by the arrows.
(b) Coordinates for the system. It is convenient to introduce two coordinates, Cartesian coordinate $\textbf{r} = (x,y)$ and
another coordinate $(X_a,Y_a)$. The coordinate $(x,y)$ is used for the region inside the junction where $x \in [0,W]$.
 $(X_a,Y_a)$ is used for the region near the boundary arc of SC $a$, where $a = \text{u}$ ($a = \text{l}$ ) for the upper (lower) SC.  $X_a \in [-l_a, 0]$ and $Y_a = 0$ along the boundary arc of the SC $a$, where $l_a$ is the length of the arc. $\theta$ is the angle between the two coordinates.
The red crosses show the positions of branch cuts representing the anti-periodic boundary condition of the modes. 
} \label{Fig:setup}
\end{figure}

In this section, we introduce the planar topological Josephson junction on a TI in Fig.~\ref{Fig:setup} and develop the theory for the junction.
In section~\ref{Subsec:Hamiltonian}, we derive the effective Hamiltonian describing the low-energy phenomena of the junction, and discuss the symmetry of the Hamiltonian. 
In section~\ref{Subsec:EmergenceMFs}, we derive the condition for the emergence of the extended chiral Majorana fermion along the arc of the SC regions outside the junction, by using a scattering matrix approach.
 In section~\ref{Subsec:FJ}, we briefly summarize the $2n\pi$ fractional Josephson effect found in Ref.~\cite{ChoiSim}.
   
%%%%%%%%%%%%%%%%%%%%%%%%%%%%%%%%%%%%% FIGURE.1 BEGIN %%%%%%%%%%%%%%%%%%%%%%%%%%%%%%%%%%%%%
%%%%%%%%%%%%%%%%%%%%%%%%%%%%%%%%%%%%% FIGURE.1 END %%%%%%%%%%%%%%%%%%%%%%%%%%%%%%%%%%%%%

\subsection{Derivation of an effective hamiltonian}\label{Subsec:Hamiltonian}
The Josephson junction in Fig.~\ref{Fig:setup} can be described by the Dirac-Bogoliubov-de Gennes Hamiltonian~\cite{FuKane08} $H_{\text{DBdG}} = \int d\textbf{r} \Psi^{\dagger}(\textbf{r}) \mathcal{H}(\textbf{r}) \Psi(\textbf{r})/2$, where 
\begin{equation}\label{Hfull}
\mathcal{H}(\textbf{r}) = \tau_z [ v_F \boldsymbol{\Pi} \cdot \boldsymbol{s} - \mu ] + \Delta_0(\textbf{r})[\tau_x \cos\varphi(\textbf{r}) + \tau_y \sin\varphi(\textbf{r})],
\end{equation}
$\Psi^{\dagger}(\textbf{r}) = \left( \psi_{\uparrow}^{\dagger}(\textbf{r}), \psi_{\downarrow}^{\dagger}(\textbf{r}), \psi_{\downarrow}(\textbf{r}), - \psi_{\uparrow}(\textbf{r}) \right)$, and $\tau_{x,y,z}$ are the Pauli matrices  in the particle-hole space.
 $\psi_{\uparrow (\downarrow)}^{\dagger}$ creates an electron with spin $\uparrow$ ($\downarrow$) at position $\textbf{r}$ on the topological insulator surface, and
$s_{x,y,z}$'s are the Pauli matrices for the electron spin. $v_F$ is the Fermi velocity of electrons on the TI surface. $\boldsymbol{\Pi}\equiv\textbf{p}-\tau_z e\boldsymbol{A}$, where 
$\textbf{p} = -i\hbar \mathbf{\nabla}$, $e<0$ is the electron charge, $\textbf{A}$ is the vector potential associated with the magnetic field $B(\textbf{r})\hat{\mathbf{z}} = \boldsymbol{\nabla}\times\textbf{A}$. $B(\textbf{r})=B$ on the surface outside the SC regions, while $B(\textbf{r}) = 0$ inside the SC regions. 
$\mu$ is the chemical potential. 
$\Delta_0( \textbf{r}) = \Delta_0$ is the magnitude of the proximity-induced superconducting order on the TI surface ($\Delta_0( \textbf{r}) = 0$ outside the SCs), and $\varphi( \textbf{r})$ is the superconducting phase of each SC region. 
The superconducting phase difference between the SCs
\begin{equation}
\Delta\varphi(x,\varphi_0) = \frac{2\pi N}{W}\left(x-\frac{W}{2}\right) - \varphi_0
\end{equation}
depends on the magnetic flux $N\equiv BLW/\Phi_0$ in the junction area $LW$, where $\varphi_0$ is the phase offset (the value in the absence of the flux), $L$ is the junction length, $W$ is the junction width, and $\Phi_0 = h/(2e)$. 

The Hamiltonian $H_\text{DBdG}$ is particle-hole symmetric $\{\Xi,H_\text{DBdG}\}=0$, where $\Xi=\tau_y\sigma_y K$, $K$ is complex conjugation, and $\Xi^2=1$. $H_\text{DBdG}$ is also invariant under $\varphi_0\mapsto\varphi_0 + 2\pi$, and the phase offset $\varphi_0$ evolves in time under a voltage bias $V(t)$ across the junction according to the Josephson relation
\begin{equation}
% V(t)= \frac{\hbar}{2e}\partial_t\varphi_0.
% \frac{\partial\varphi_0}{\partial t} = \frac{2e}{\hbar}V(t).
% \partial_t \varphi_0(t) = \frac{2e}{\hbar}V(t).
\varphi_0(t) = \frac{2e}{\hbar}\int_0^tV(t')dt'. 
\label{Eq:JosephsonRelation}
\end{equation} 
%We discuss the dynamics of the Josephson junction under various $V(t)$ in the following sections.

%Around the boundary of the topological SC region $a$ ($a = \textrm{u}$ for the upper SC and $a= \textrm{l}$ for the lower SC) of coordinate $(X_a,Y_a)$ [see Fig.~\ref{SuppFig1}], the vector potential can be approximated into $\textbf{A}(X_a,Y_a) = -B(Y_a + s_a L/2,0,0)$ under the condition $\l_B \ll R$, where  $s_\textrm{u} = 1$, $s_\textrm{l} = -1$, $L$ is the junction length, and $1/R$ is the curvature of the boundary.  

%Assuming that we have turned on the magnetic field slow enough for the initial state to reach for its ground state, $\varphi_0 = \pi$ for the odd number of $N$. As turning on the voltage bias, $\varphi_0 \rightarrow  \varphi_0 + \frac{2eV_{DC}}{\hbar}t$ so that $\Delta\varphi(\mathbf{r},t) = \pi + \frac{2eV_{DC}}{\hbar}t - \frac{2\pi N x}{W}$.  We mention that during $T=h/(2eV_{DC})$, the difference of superconducting phase evolves by $2\pi$. 

 %In the absence of the magnetic field, $\varphi( \textbf{r}) = 0$ in the lower SC, while $\varphi(\textbf{r}) = \varphi$ in the upper SC. 

We derive a low-energy effective Hamiltonian $H$  from the full Hamiltonian $H_{\text{DBdG}}$, focusing on the energy scale $\lesssim \Delta_0$. 
We first discuss the limit of $L \to \infty$, where the two SC regions have no wave-function overlap and are independent.
In this limit, chiral modes $\eta_\textrm{u(l)}$ propagate along the boundary between the upper (lower) SC region and the gapped TI region (see the yellow lines in Fig.~\ref{Fig:setup}). The chiral modes appear, since the SC regions are topological while the gapped TI region is topologically trivial, according to the bulk-edge correspondence.
The chiral propagating modes are described by the kinetic Hamiltonian of
$H_{L \to \infty} = -i\hbar v_{\text{arc}} \int_{-l_{\textrm{u}}}^{W}\eta_{\textrm{u}}(X_{\textrm{u}}) \partial_{X_{\textrm{u}}} \eta_{\textrm{u}}(X_{\textrm{u}}) dX_{\textrm{u}} + i\hbar v_{\text{arc}} \int_{-l_{\textrm{l}}}^{W}\eta_{\textrm{l}}(X_{\textrm{l}}) \partial_{X_{\textrm{l}}} \eta_{\textrm{l}}(X_{\textrm{l}}) dX_{\textrm{l}}$, where $X_\textrm{u(l)} \in [-l_\textrm{u(l)}, W]$ is the coordinate along the boundary of the upper (lower) SC region and $v_\textrm{arc}$ is the propagativing velocity.
%In the limiting case, two chiral zero-energy modes of $\eta_\textrm{u}$ and $\eta_\textrm{l}$ exist; $\eta_\textrm{u}$ surrounds along the boundary of the upper SC, and another mode $\eta_\textrm{l}$ does along the boundary of the lower SC. 
The operators $\eta_\textrm{u(l)}$ of the chiral modes are written in terms of the Nambu spinor $\Psi (\textbf{r})$ of the full Hamiltonian $H_{\text{DBdG}}$ as $ \eta_a (X_a) = \int \Psi^{\dagger}(X_a,Y_a)\Phi_a (X_a,Y_a)dY_a$ at a point  $X_{a = \textrm{u,l}}$ on the boundary, where $\Phi_a (X_a,Y_a)$ is the wave function of the chiral mode and satisfies  $\Xi\Phi_a(X_a,Y_a)=\Phi_a(X_a,Y_a)$.
%at a point $X_a$ on the boundary arc of SC $a$, where $X_a\in(-l_a,W)$, $a = \textrm{u}$ for the upper SC and $a = \textrm{l}$ for the lower SC, and $\Xi\Phi_a(X_a,Y_a)=\Phi_a(X_a,Y_a)$.  
Because of the particle-hole symmetry $\Xi$, the chiral modes have the redundant degree of freedom in creation and annihilation,
\begin{equation}
\eta_a^\dagger(X_a) = \eta_a(X_a).
\end{equation}

The chiral modes satisfy the nontrivial boundary condition of 
\begin{equation}
\eta_a(x=W) = -(-1)^{M_a}\eta_a(x=-l_a), \label{Eq:BC}
\end{equation}
where $M_a$ is the number of Abrikosov vortices in the upper (lower) SC region of $a=\text{u}$ ($a=\text{l}$). The condition has the two sign factors $(-1)$ and $(-1)^{M_a}$. The first sign factor $(-1)$ originates from the Berry phase $\pi$ due to the momentum-spin locking of the TI surface; the chiral modes gain the Berry phase while they circulate once along the boundary; as a result, the wave function $\Phi_a (X_a,Y_a)$ satisfies the anti-periodic boundary condition in the absence of the Abrikosov vortices. The second factor $(-1)^{M_a}$ is due to the fact that
each Abrikosov vortex hosts a localized Majorana zero mode and results in $\pi$ phase that the wave function $\Phi_a(X_a,Y_a)$ gains while it winds around the vortex.~\cite{Akhmerov, Law}

%  Moreover, the presence of the Abrikosov vortices in SCs, which bind localized Majorana fermions, changes the boundary condition of $\Phi_a(X_a,Y_a)$ due to the superconducting phase winding around each vortice~\cite{Akhmerov, Law}. Accordingly, the boundary condition of the chiral modes is 

%$\eta_a(x=W) = -(-1)^{M_a}\eta_a(x=-l_a)$ depending on the number of the Abrikosov vortices $M_a$ in SC region of $a$.

We next consider the short-junction regime of finite $L \ll \xi$, 
where $\xi\equiv\hbar v_F/\Delta_0$ is the SC coherence length. To obtain the effective Hamiltonian $H$, we project the full Hamiltonian  $H_\text{DBdG}$ onto the state space of the chiral modes in the same spirit as in the envelope-function theory.~\cite{EFtheory} In this approach, the chiral modes are chosen as the basis states.
The two SC regions now have wave-function overlap, hence, tunneling between the two chiral modes $\eta_\textrm{u}$ and $\eta_\textrm{l}$ happens.
As a result, the Hamiltonian $H$ is decomposed into the junction part $H_\textrm{J}$ and the arc part $H_\textrm{arc}$ as $H = H_\textrm{J}+ H_\textrm{arc}$,
%We first obtain the zero-energy modes of $H_{\text{DBdG}}$ in the limiting case of $L = 0$ and $\Delta \varphi = \pi$, and we project $H_{\text{DBdG}}$ onto the space of the zero-energy modes.
%Now, we project the $H_\text{DBdG}$ onto the above chiral zero-energy modes $\eta_a$ and obtain the effective Hamiltonian $H=H_{\text{arc}}+H_{\text{J}}$ containing kinetic and tunneling energy of the chiral modes $\eta_a$, where $H_{\text{arc}}$ and $H_{\text{J}}$ describes the region of the boundary arcs of SCs and junction, respectively.
\begin{widetext}
\begin{eqnarray}
H_{\text{arc}} &=& -i\hbar v_{\text{arc}} \int_{-l_{\textrm{u}}}^{0}\eta_{\textrm{u}}(X_{\textrm{u}}) \partial_{X_{\textrm{u}}} \eta_{\textrm{u}}(X_{\textrm{u}}) dX_{\textrm{u}} + i\hbar v_{\text{arc}} \int_{-l_{\textrm{l}}}^{0}\eta_{\textrm{l}}(X_{\textrm{l}}) \partial_{X_{\textrm{l}}} \eta_{\textrm{l}}(X_{\textrm{l}}) dX_{\textrm{l}}  \\
H_{\text{J}} &=& -i\hbar v_{\text{J}} \int_{0}^{W}\eta_{\textrm{u}}(x) \partial_{x} \eta_{\textrm{u}}(x) dx + i\hbar v_{\text{J}} \int_{0}^{W}\eta_{\textrm{l}}(x) \partial_{x} \eta_{\textrm{l}}(x) dx \nonumber \\
&& -i \int_{0}^{W} \eta_{\textrm{u}}(x) m(x,\varphi_0) \eta_{\textrm{l}}(x) dx +i \int_{0}^{W} \eta_{\textrm{l}}(x) m(x,\varphi_0) \eta_{\textrm{u}}(x) dx.
\end{eqnarray}
\end{widetext}
The junction-part Hamiltonian $H_\textrm{J}$ contains the term for the coupling $m(x,\varphi_0)$ between the chiral modes in addition to the kinetic term. The coupling depends on the position-dependent SC phase difference $\Delta \varphi$ across the junction as
\begin{equation}
m(x,\varphi_0) = \frac{\Delta_0 }{1+ L/\xi}\cos\frac{\Delta\varphi(x,\varphi_0)}{2}.
\end{equation}
%The low-energy effective Hamiltonian $H$ is valid, when the SC phase difference across the junction spatially varies much slower than the wave functions of the chiral modes. 
The expression of the wave function $\Phi_a(X_a,Y_a)$ of the chiral modes in this short-junction regime is found in Appendix~A.

The propagating velocity $v_\text{J}$ of the chiral modes inside the junction is different from the velocity $v_\text{arc}$ along the SC arcs outside the junction. The difference can be understood from the following facts.
The propagation inside the junction occurs with the specular or retro Andreev reflections between the two SC regions.
By contrast, the propagation along the arc is determined by skipping trajectories due to the magnetic field.
We derive the velocities 
\begin{eqnarray*}
&& \frac{v_{\text{arc}}}{v_F} = \frac{6 \tilde{\mu}  \Gamma \left(-\frac{\tilde{\mu} ^2}{4}\right) \Gamma \left(1-\frac{\tilde{\mu} ^2}{4}\right)-6 \tilde{\mu}  \Gamma \left(\frac{1}{2}-\frac{\tilde{\mu} ^2}{4}\right)^2}{ \left(\tilde{\mu} ^2 \left(\psi \left(-\frac{\tilde{\mu} ^2}{4}\right)-\psi \left(\frac{1}{2}-\frac{\tilde{\mu} ^2}{4}\right)\right)-2\right)D},  \\ 
&& \frac{v_{\text{J}}}{v_F} = \frac{ \cos \frac{\mu L }{\hbar v_F} + \frac{\Delta_0}{\mu}\sin\frac{\mu L}{\hbar v_F}}{1+ L/\xi}\frac{\Delta_0^2}{\Delta_0^2 + \mu^2},  \label{velocityJv} 
\end{eqnarray*}
where $\tilde{\mu} \equiv \mu l_B/(\hbar v_F)$, $l_B=\sqrt{\hbar/(eB)}$ is the magnetic length, $\Gamma$ and $\psi$ are the Gamma function and Digamma function, respectively, and $D\equiv \Gamma \left(-\frac{\tilde{\mu} ^2}{4}\right) \Gamma \left(\frac{1}{2}-\frac{\tilde{\mu} ^2}{4}\right)$. Depending on the chemical potential $\mu$, $v_\textrm{arc}$ can be much larger (one or two order larger) than $v_\textrm{J}$.

In the case that the upper and lower SC regions have the same arc length of $l_\textrm{u} = l_\textrm{l}=l$, we write the low-energy effective Hamiltonian $H$ in the simpler form of
\begin{equation}
H = \int_{-l}^{W}\Gamma^\top(x)\mathcal{H}(x)\Gamma(x)dx, \quad \Gamma^{\top}(x)\equiv\left(\eta_\text{u}(x), \,\, \eta_\text{l}(x)\right),  \label{Eq:Heff}
\end{equation}
merging the coordinates $x$ and $X_a$ so that $x$ extends as  $x \in [-l, W]$.
Here  $\mathcal{H}(x)=[-i\hbar v_\text{J}\partial_x\sigma_z]\Theta(x) + [-i\hbar v_\text{arc}\partial_x\sigma_z + m(x,\varphi_0)\sigma_y]\Theta(-x)$ and $\Theta(x)$ is the Heaviside function. 
In the rest of paper, we use this form of the low-energy effective Hamiltonian. 
Note that this Hamiltonian in Eq.~\eqref{Eq:Heff} was used in Ref.~\cite{ChoiSim}.
  
%$\Delta \varphi = \pi$

We discuss the symmetries of the low-energy effective Hamiltonian in Eq.~\eqref{Eq:Heff}.
Inheriting the invariance of the parent Hamiltonian $H_\text{DBdG}$, 
$H$ is particle hole symmetric $\{H,K\}=0$ and invariant under $\varphi_0\mapsto\varphi_0+2\pi$.
The invariance of $H$ under $\varphi_0\mapsto\varphi_0+2\pi$ is achieved, as $m(x,\varphi_0+2\pi)=-m(x,\varphi_0)$, $\eta_\text{u}\rightarrow-\eta_\text{u}$, and $\eta_\text{l}\rightarrow\eta_\text{l}$.
% In addition, resembling the Su-Scheieffer-Heeger model,~\cite{FuKane08} $H$ has the chiral symmetry of $\{H,\sigma_x\}=0$ that is valid only when $l_\textrm{u} = l_\textrm{l}=l$.
   
%$L/\xi \ll 1$ and the typical case of $\Delta_0 / \mu \ll 1$, we find $v_{\text{J}} \sim (\Delta_0/\mu)^2v_F$, and $m(x,\varphi_0) = \Delta_0 \cos(N\pi x/W-\varphi_0/2)$/. Moreover, under a DC voltage bias $V_\text{DC}$, $m(x,t) = \Delta_0 \sin(N\pi x/W-eV_\text{DC}t/\hbar)$, where we choose time $t=0$ when $\Delta\varphi(x=0,t) = \pi$. 

%We discuss the case of $l_\textrm{l} \ne l_\textrm{u}$. For this case, the quantization condition in Eq. (2) in the main text remains as the same form but with the replacement of $2l$ by $l_\textrm{l} + l_\textrm{u}$, as $2l$ is the total length of the boundary arcs of the upper and lower SCs. This modification does not affect the zero-energy states of $E=0$. Hence, the case of $l_\textrm{l} \ne l_\textrm{u}$ has the qualitatively same results as the case of  $l_\textrm{u} = l_\textrm{l}$.

\subsection{Majorana zero modes of the system}\label{Subsec:EmergenceMFs}

The Hamiltonian in Eq.~\eqref{Eq:Heff} has two different types of Majorana zero modes, localized ones inside the junction and extended ones along the arcs outside the junction, in addition to the other midgap states.
The localized Majorana zero modes have been discussed in Refs.~\cite{PotterFu}. The extended chiral Majorana zero mode arising due to the  finite-size SC regions found in Ref.~\cite{ChoiSim}.
We below discuss when the Majorana zero modes occur.

The localized Majorana zero modes $\gamma_{k}$ are expressed as $\gamma_{k} = \int \Gamma^{\top}(x)\psi_{k}(x) dx$ with the wave function $\psi_{k}(x)$ localized around $x_k$ at which the sign of $m(x,\varphi_0)$ changes~\cite{FuKane08},  
\begin{equation}
\psi_k(x) = \sqrt{\frac{N}{4WI_0(2\delta)}}e^{\delta\cos\frac{N\pi}{W}(x-x_k)}
\left(
\begin{array}{c}
1 \\
-(-1)^k
\end{array}
\right). \label{Eq:wavefunctionMF}
\end{equation}
Here $\delta \equiv (\frac{ W }{ \pi N \lambda_M})^2$, $\lambda_M=\sqrt{\hbar v_\text{J}W/(\pi N\Delta_0)}$ is the localization length of the zero mode, and $I_0(x)$ is the modified Bessel function of the first kind. The localization center is located at
\begin{equation}
x_k=\frac{W}{N} \left( \frac{2k-1}{2}+\frac{\varphi_0}{2\pi} \right)  + \frac{W}{2}, \label{Eq:positionMF}
\end{equation}
where $0\le x_k\le W$ and $k$'s are integers.
The distance between the adjacent localized Majorana zero modes is $W/N$, and there are up to $N$ localized Majorana zero modes inside the junction.

The condition of the existence of the localized Majorana zero modes is
\begin{equation}
\delta =  \left( \frac{ W }{ \pi N \lambda_M} \right)^2 \gg 1, 
\end{equation}
as $\sqrt{\delta}$ is the ratio of the mode separation to the localization length.
In this case of $\delta \gg 1$, we estimate the energy splitting of the Majorana zero modes due to the overlap of adjacent zero modes as $|\int \psi_k^\top(x)\mathcal{H}(x) \psi_{k+1}(x) dx|\approx \Delta_0 e^{-\pi^2\delta/4}$, which is indeed negligible at $\delta \gg 1$.
Note that the wave-function form of $\psi_k(x)$ in the limit of $\delta \gg 1$ was found in  Ref.~\cite{twoMFs}.

We emphasize that the localized Majorana zero modes move in the junction when a DC voltage $V_\text{DC}$ bias is applied across the junction. In this case, the localization centers $x_k$ in Eq.~\eqref{Eq:positionMF} become time dependent according to the Josephson relation in Eq.~\eqref{Eq:JosephsonRelation}. This fact was used in Ref.~\cite{ChoiSim}, to move and braid Majorana fermions.
 
Next we discuss the condition when the extended chiral Majorana zero mode propagating along the arcs occurs. 
We derive the condition by using a scattering matrix for the junction region.
For example, the chiral mode of energy $E$ approaches from the arc to the left end of the junction. This mode is reflected to the chiral mode at the left end with reflection amplitude $r_E$ or transmitted to the chiral mode at the right end of the junction with transmission amplitude $t_E$. Similarly, the chiral mode approaching from the arc to the right end of the junction is reflected at the same end with reflection amplitude $r_E'$ or transmitted to the opposite end with transmission amplitude $t_E'$.
These are described by the scattering matrix
\begin{equation}\label{Smatrix}
\left(
\begin{array}{cc}
r_E & t'_E \\
t_E & r'_E
\end{array}
\right)
\left(
\begin{array}{c}
A\\
B
\end{array}
\right)_E
=
\left(
\begin{array}{c}
C\\
D
\end{array}
\right)_E,
\end{equation}
where $A$ and $B$ ($C$ and $D$) are the wave function amplitudes of the chiral mode incoming to (outgoing from) the left and right ends of the junction, respectively.
 %We define two amplitudes $A$ and $B$ of incident chiral modes into left and right end of the junction, respectively, while defining $C$ and $D$ as amplitudes of chiral modes reflected from left and right end of the junction, respectively. The scattering matrix of the junction connects the incident and reflected amplitudes at energy $E$ as

The wave function amplitude $C$ of the chiral mode outgoing from the left end of the junction is related with the amplitude $B$ of the chiral mode incoming to the right end as
\begin{equation}
B=- Ce^{iEl_\text{l}/(\hbar v_\text{arc})+i\pi M_\text{l}}, \label{BCrelation}
\end{equation}
since the outgoing mode becomes the incoming mode after propagation along the lower arc of length $l_\text{l}$. 
The sign factor $-1$ originates from the Berry phase of the chiral mode $\eta_\text{l}$,
$El_1 / (\hbar v_\textrm{arc})$ is the dynamical phase gain,
and $\pi M_\textrm{l}$ counts the phase due to the vortices in the lower SC (whose number is $M_\textrm{l}$) as in Eq.~\ref{Eq:BC}.
Similarly, $D$ and $A$ are related as
\begin{equation}
A=- D e^{iEl_\text{u}/(\hbar v_\text{arc})+i\pi M_\text{u}}. \label{ADrelation}
\end{equation}

Combining Eq.~\eqref{Smatrix},  \eqref{BCrelation}, and \eqref{ADrelation} and requiring nonzero solutions of $A$ and $B$, we find the quantization condition for the chiral mode propagating along the arcs at finite energy $E \to 0$
%\begin{equation}
%\left( \begin{array}{cc}
%r_E & t'_E + e^{-iEl_\text{l}/v_\text{arc}+i\pi M_\text{l}}\\
%t_E + e^{-iEl_\text{u}/v_\text{arc}+i\pi M_\text{u}} & r'_E
%\end{array} \right).
%\end{equation}
%$r_Er'_Ee^{iE(l_\text{u}+l_\text{l})/v_\text{arc}+i\pi(M_\text{u}+M_\text{l})}=1$.
\begin{equation}
r_Er'_Ee^{iE(l_\text{u}+l_\text{l})/(\hbar v_\text{arc})+i\pi(M_\text{u}+M_\text{l})}=1. \label{Eq:QuantRule2}
\end{equation}
%We derive analytic expressions of the scattering matrix with focus on the finite energy and limiting case $E\rightarrow 0$, since Majorana fermions are not bound at the exact zero-energy but with exponentially small energy splitting from $E=0$.   We note that $t_E, t'_E$ immediately vanish with finite $E$ because the wave function in the junction starts to decay exponentially with finite energy, resulting in $|t_E|^2, |t'_E|^2 \le 1/(1+4(E/\Delta_0)^2\sinh^4 \delta)$. For typical parameters giving localized Majorana fermions in the junction, we estimate $\delta\sim 9$, and the energy at which $|t_E|^2, |t'_E|^2 = 1/2$ is order of $10^{-8}\,\text{meV}$, which corresponds to the temperature of $10^{-4}\,\text{mK}$. 
%The wave function penetrates through the junction only at the exact zero-energy. 
Approximating $m(x,\varphi_0)$ as a linear function of $x$ at $x = 0$ and $x=W$, we derive
\begin{eqnarray}
r_E
&=&\frac{D_{\nu}\left(z\right) + i\frac{E}{\Delta_0}\sqrt{\frac{\delta}{2|\sin\frac{\varphi_0}{2}|}}D_{\nu-1}(z)}{D_{\nu}(z) - i\frac{E}{\Delta_0}\sqrt{\frac{\delta}{2|\sin\frac{\varphi_0}{2}|}}D_{\nu-1}(z)}\\
r'_E
&=&-\frac{D_{\nu'}\left(z'\right) + i\frac{E}{\Delta_0}\sqrt{\frac{\delta}{2|\sin\frac{\varphi_0}{2}|}}D_{\nu'-1}(z')}{D_{\nu'}(z') - i\frac{E}{\Delta_0}\sqrt{\frac{\delta}{2|\sin\frac{\varphi_0}{2}|}}D_{\nu'-1}(z')}
\end{eqnarray}
where $\nu\equiv\frac{E^2\delta}{2\Delta_0^2|\sin\frac{\varphi_0}{2}|}$, 
$z\equiv\frac{\sqrt{2}\delta}{\sqrt{|\sin\frac{\varphi_0}{2}|}}\cos\frac{\varphi_0}{2}$, 
$\nu'\equiv\frac{E^2\delta}{2\Delta_0^2|\sin\left(\pi N-\frac{\varphi_0}{2}\right)|}$, $z\equiv\frac{\sqrt{2}\delta}{\sqrt{|\sin\left(\pi N-\frac{\varphi_0}{2}\right)|}}\cos\left(\pi N-\frac{\varphi_0}{2}\right)$, and $D_{\nu}(z), D_{\nu'}(z')$ are the parabolic cylinder functions. We find $r_{E\rightarrow0}=\text{sgn}[m(x=0,\varphi_0)]$ and $r'_{E\rightarrow0}=-\text{sgn}[m(x=W,\varphi_0)]$ when the localized Majorana zero modes of the junction are spatially well-separated from the two ends of the junction. %, while the reflection phases of $r_E$ and $r'_E$ continuously change between $0$ and $\pi$ as localized Majorana fermions are pulled out of the end of the junction. 
We find that the analytic results of $r_E$ and $r'_E$ are in good agreement with the numerical result obtained without the approximation of $m(x, \varphi_0)$.

Using the analytic expressions of  $r_E$ and $r'_E$, we write the quantization condition of the extended chiral mode in Eq.~\eqref{Eq:QuantRule2} in a comprehensible form,
\begin{equation}
E(l_\text{u}+l_\text{l})/(\hbar v_\text{arc}) + \pi + \pi(M_M + M_\text{u}+M_\text{l})= 0, 2\pi, \cdots. \label{Eq:QuantRule}
\end{equation}
Here $M_M$ is the number of sign changes in $m(x,\varphi_0)$, which equals to the number of localized Majorana zero modes in the junction. This quantization condition shows that
at  $E\rightarrow0$, the extended chiral Majorana zero mode  appears (disappears) along the arcs when the total number of localized Majorana zero modes in the Josephson and at Abrikosov vortices is odd (even).  This is in good agreement with the fact that Majorana zero modes always come in pairs in electron systems. This is indeed independent of the specific shapes of the SC regions. In section~\ref{Sec:MultipleMFs}, we exemplify the emergence of Majorana zero modes in our setup for the various number of the localized Majorana zero modes in the junction and the
various number of the Abrikosov vortices in the SC regions.

%%%%%%%%%%%%%%%%%%%%%%%%%%%%%%%%%%%%% FIGURE BEGIN %%%%%%%%%%%%%%%%%%%%%%%%%%%%%%%%%%%%%
\begin{figure*}[ht]
%\centering
\includegraphics[width=1.5\columnwidth]{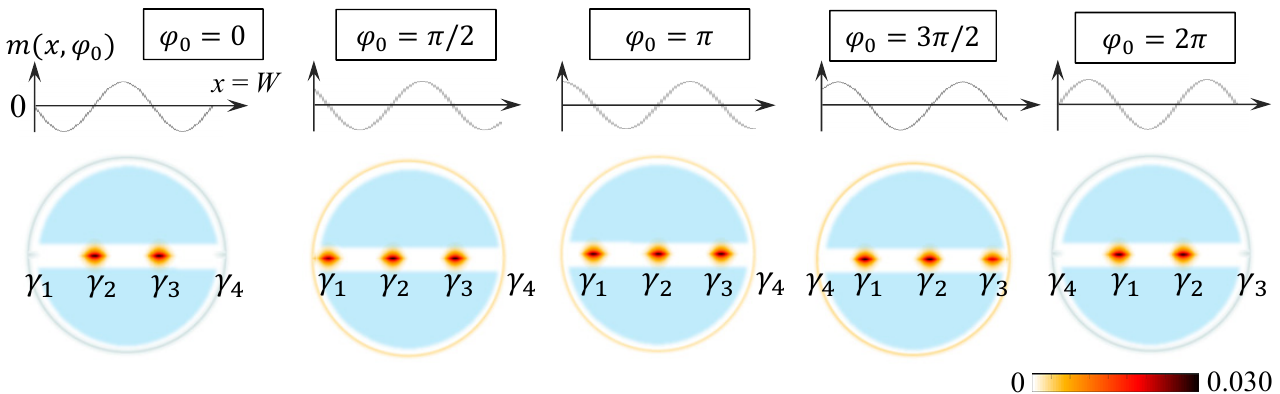}
\caption{For  $N=3$ and $M_\text{u}=M_\text{l}=0$,  the spatial dependence of the coupling strength $m(x, \varphi_0)$ (top pannels) and the probability density of Majorana fermions $\gamma_{k=1,2,3,4}$ (bottom panels) are shown for various $\varphi_0$. 
When $\gamma_k$'s are Majorana zero modes, their probability is shown in color (see the scale bar). When two of $\gamma_k$'s are fused, the probability is shown in grey.
The SC regions are represented in blue for a guide. Notice that the spatial distribution of the probability density is $2\pi$ periodic in $\varphi_0$. The numerical results are obtained with $\Delta_0=1\,\text{meV}$, $v_F=5.0\times10^{5}\,\text{m/s}$, $v_\text{J}=0.04v_F$, $v_\text{arc}\sim v_F$, $W=800\,\text{nm}$, and $l_\text{u}=l_\text{l}=1.26\,\mu\text{m}$.
} \label{Fig:MFsN3}
\end{figure*}
%%%%%%%%%%%%%%%%%%%%%%%%%%%%%%%%%%%%% FIGURE END %%%%%%%%%%%%%%%%%%%%%%%%%%%%%%%%%%%%%
%%%%%%%%%%%%%%%%%%%%%%%%%%%%%%%%%%%%%% FIGURE BEGIN %%%%%%%%%%%%%%%%%%%%%%%%%%%%%%%%%%%%%
%\onecolumngrid
%\begin{center}
%\begin{figure}[pt]
%%\centering
%\includegraphics[width=0.85\linewidth]{prbMFsN3.pdf}
%\caption{(Color online) Energy spectrum is depicted as a function of $\varphi_0$ under $N=3$ and $M_\text{u}=M_\text{l}=0$ in (a). The spatial probability density of the energy levels which is closest to the zero-energy are drawn for various $\varphi_0$ in (b) along with $m(x,\varphi_0)$ on the top. Majorana fermions are coloured with the scale at the bottom right, while the fused Majorana fermions are depicted with grey. SC regions are shaded with blue for a guide. Notice that energy levels and probability densities are $2\pi$ periodic in $\varphi_0$. We use parameters as follows. $\Delta_0=1\,\text{meV}$, $v_F=5.0\times10^{5}\,\text{m/s}$, $v_\text{J}=0.04v_F$, $v_\text{arc}\sim v_F$, $W=800\,\text{nm}$, and $l_\text{u}=l_\text{l}=1.26\,\mu\text{m}$.
%} \label{Fig:MFsN3}
%\end{figure}
%\end{center}
%\twocolumngrid
%%%%%%%%%%%%%%%%%%%%%%%%%%%%%%%%%%%%%% FIGURE END %%%%%%%%%%%%%%%%%%%%%%%%%%%%%%%%%%%%%

%$\bullet${\color{red} adiabatic regime of $T_J \gg \hbar/E_0$}
%  
%$\bullet${\color{red} period-elongation factor $n$}

\subsection{Short review of $2n\pi$ fractional AC Josephson effect}\label{Subsec:FJ}

In Ref.~\cite{ChoiSim}, it was shown that the junction can have four Majorana zero modes $\gamma_{k=1,2,3,4}$ when $N=3$ and in the absence of Abrikosov vortices ($M_\textrm{u} = M_\textrm{l} = 0$). When a weak DC voltage bias $V_\textrm{DC}$ is applied across the junction, they move along the junction adiabatically. While moving, the interchange of fusion partners of the four Majorana zero modes happens, resulting in non-Abelian state evolution and the $2n \pi$ fractional AC Josephson effect.
In this section, we briefly review the mechanism of the non-Abelian state evolution and the $2n\pi$ fractional AC Josephson effect, since it is a background of this work.

%The non-Abelian effect of Majorana zero modes have been discussed in the finite SC-TI-SC Josephson junction with $N=3$ and $M_\text{u}=M_\text{l}=0$ in Ref.~\cite{ChoiSim}. The Majorana zero modes are split and braided adiabatically $\hbar/\Delta_0 \ll T_J$. Finally, they fuse again as interchanging their fusion pair, which causes the $2n\pi$ fractional AC Josephson effect~\cite{ChoiSim}.

%We note that $\varphi_0$ evolves in time according to Eq.~\eqref{Eq:JosephsonRelation} as $\varphi_0(t)= 2eV_\text{DC}t/\hbar$. We focus on the even fermion parity states. 

The junction has the two lowest single-particle energy levels.
We focus on the four Majorana fermions $\gamma_{k=1,2,3,4}$  associated with the two energy levels.
Figure~\ref{Fig:MFsN3} shows the dependence of the spatial distribution of the Majorana fermions $\gamma_{k=1,2,3,4}$ on the SC phase difference $\varphi_0$ and how the interchange of fusion partners of the four Majorana zero modes happens.
At each $\varphi_0$, the positions of the localized Marajona fermions are shown in Fig.~\ref{Fig:MFsN3}, and they follow Eq.~\eqref{Eq:positionMF}. The localization length is much shorter than their separation.
At $\varphi_0 = 0$ and $2\pi$, the two Majorana fermions, one at the left junction end and the other at the right end, fuse along the arcs of the SC region.
At $\varphi_0 = \pi$, three Majorana fermions of the four are localized Majorana zero modes inside the junction, while the other one is formed along the arcs as the extended chiral Majorana zero mode; the formation of the extended mode follows Eq.~\eqref{Eq:QuantRule}. In this case, the four Majorana zero modes form two-fold degenerated ground states.

As $\varphi_0$ is continuously increased by the DC voltage $V_\textrm{DC}$, the four Majorana fermions evolve, following the panels of Fig.~\ref{Fig:MFsN3}.
Although the positions of the four Majorana fermions are $2\pi$-periodic in $\varphi_0$, the fusion partners are different between $\varphi_0 = 0$ and $2\pi$.
At $\varphi_0 = 0$, $\gamma_1$ and $\gamma_4$ fuse along the arcs.
By contrast, $\gamma_3$ and $\gamma_4$ fuse along the arcs at $\varphi_0 = 2\pi$.
Namely the interchange of the fusion partners happens as
$\{ (\gamma_4, \gamma_1), (\gamma_3, \gamma_2) \}_{\varphi_0 =0} \to (\gamma_4, \gamma_3), (\gamma_2, \gamma_1) \}_{\varphi_0 =2\pi} \}$. 
Similarly, the wave functions of the Majorana fermions at $\varphi_0 =0$
are related with those at $\varphi_0 = 2\pi$ as~\cite{ChoiSim}
\begin{equation}
\gamma_{i=1,2,3}(\varphi_0 = 2\pi)=\gamma_{i+1}(0), \quad \gamma_4(2\pi) = -\gamma_1(0).	\label{Eq:braiding}
\end{equation}
%  Initially, $\varphi_0 = 0$ minimizing the ground state energy [see Fig.~\ref{Fig:MFsN3}(a)]. The phase advancing of $\varphi_0$ splits the pair of Majorana fermions $\gamma_1$ and $\gamma_4$ into zero modes, and four Majorana fermions are fully split when $\varphi_0=\pi$ so that two-fold degenerated ground states are obtained. As $\varphi_0$ advances further, the train of localized Majorana fermions $\gamma_1,\gamma_2,\gamma_3$ moves to the right end of the junction, and the fusion pair of Majorana fermions interchanges into $\gamma_3$ and $\gamma_4$. 
As a result, the time-evolution of the two complex fermion states defined by the four Majorana fermions after one conventional Josephson period $T_\textrm{J} = h / (2e V_\textrm{DC})$ (corresponding to the $2\pi$ change of $\varphi_0$) is described by the unitary matrix $U$
\begin{equation}
U=U_{\phi'}U_B U_{\phi}=\frac{1}{\sqrt{2}}\left(
\begin{array}{cc}
e^{i\phi_+} & e^{-i\phi_-} \\
e^{i\phi_-} & -e^{-i\phi_+}
\end{array}
\right).	\label{Eq:Ue}
\end{equation}
Here, the matrix is written in the basis  $\{|0_{41}0_{32}\rangle_0, |1_{41}1_{32}\rangle_0\}$
associated with fermion operators $f_{41}$ and $f_{32}$, where $f_{ij}=(\gamma_i(0)+i\gamma_j(0))/2$ is formed by two Majorana fermions $\gamma_i$ and $\gamma_j$, 
$f_{41} |0_{41}0_{32}\rangle_0 = f_{32}|0_{41}0_{32}\rangle_0 = 0$, and $|1_{41}1_{32}\rangle_0=(f_{41})^\dagger(f_{32})^\dagger|0_{41}0_{32}\rangle_0$.
Note that only the even-parity basis states are discussed here.
The phases $\phi_\pm$ are related with the dynamical phase accumulated during the fusion and splitting of the Majorana fermions. 

As a result of the interchange of the fusion partners of the four Majorana fermions, the $2n \pi$ fractional AC Josephson effect occurs.
The period-elongation factor $n$ of the $2n \pi$ fractional effect is determined by $\phi_+$.
The evolution matrix $U$ in Eq.~\eqref{Eq:Ue} rotates an initial state such as $|0_{41}0_{32}\rangle_0$ on the Bloch sphere for the basis states  $\{|0_{41}0_{32}\rangle_0, |1_{41}1_{32}\rangle_0\}$
about the axis $\hat{r}=(\cos\phi_-\sin\theta, \sin\phi_-\sin\theta, \cos\theta)$ by the rotation angle $\Omega=2\arccos\left(\frac{\sin\phi_+}{\sqrt{2}}\right)$, where $\theta = \arctan\frac{1}{\cos\phi_+}$. The evolution recovers its initial state after $n$-th rotations (after $nT_\text{J}$) if $\Omega/n$ is an integer multiple of $2\pi$. Therefore, the period of the Josephson junction is elongated into $nT_\text{J}$. The period-elongation factor $n$ is larger than 2, and can be tuned
by $V_\text{DC}$, since  $\phi_+ \propto 1/ V_\textrm{DC}$.
This is a unique feature of the $2n \pi$ fractional AC Josephson effect, stemming from the non-Abelian fusion rule.~\cite{Nayak,Stern} The contribution  $I_\text{MF}$ of the four Majorana fermions $\gamma_k$'s  to the Josephson current is $nT_\text{J}$ periodic; 
the contribution from the midgap states is $2\pi$ periodic, as shown later in this work.

% We point out that Eq.~\eqref{Eq:Ue} includes the tunable phase of $\phi_+$, which is overlooked in Ref.~\cite{FuKane08} using a number of Josephson junctions. 

% This also contrasts to the evolution of the current-carrying states by the fermion parity switching~\cite{Kitaev1D,FuKanePRB} in which $U_B$ is replaced into $\sigma_x$ and $U_e=U_\phi\sigma_x U_{\phi'}$ satisfies $U_e^2=1$ regardless of $\phi_\pm$.

We note that $I_\text{MF}(t)$ carries the information of the state  $|Q(t)\rangle=\alpha(t)|0_{41}0_{32}\rangle_0 +\beta(t)|1_{41}1_{32}\rangle_0$ formed by the four Majorana fermions at time $t$. 
The basis states $|0_{41}0_{32}\rangle_0$ and $|1_{41}1_{32}\rangle_0$ are the partners of the particle-hole symmetry ($\Xi|0_{41}0_{32}\rangle_0=|1_{41}1_{32}\rangle_0$) and contribute to the supercurrent in the oppsite direction as $\Xi\hat{J}(x)\Xi^{-1}=-\hat{J}(x)$, where $\hat{J}(x)$ is the current density operator (see Appendex~B).
Therefore, $I_\text{MF}(t) \propto |\alpha(t)|^2 - |\beta(t)|^2$; each charge-transfer event across the junction occurs stochastically by $|0_{41}0_{32}\rangle_0$ or $|1_{41}1_{32}\rangle_0$ with probability $|\alpha|^2$ or $|\beta|^2$, respectively, and  $I_\text{MF}(t) \propto |\alpha(t)|^2 - |\beta(t)|^2$ is measured experimentally by taking average of those events during a sampling time. 

Note that the time resolved detection of the Josephson current requires the average of the current over a sampling time window which is wide enough to collect sufficient data but much shorter than the conventional Josephson period $T_\text{J}$. It is possible to have a sufficiently wide sampling window. We estimate a typical sampling window as 0.03 ns during which a hundred of the charge-transfer events can occur across the junction.~\cite{BTK} This sampling time window is much shorter than $T_\text{J}$ which is typically a few nanoseconds. 

%{\color{red} Moreover, the local fermion parity of the current-carrying state of two fusing Majorana fermions is conserved as long as the distance between localized Majorana fermion $W/N$ is larger than the size of the Cooper pair $\xi$, since transitions between $|00\rangle$ and $|11\rangle$ by Cooper pair tunneling are suppressed in the regime.}

We derive the contribution $I_\textrm{MF}$ to the Josephson current from the four Majorana fermions $\gamma_{k=1,2,3,4}$.
The contribution is related with the energy $E_\text{MF}$ of the complex-fermion state formed by the Majorana fermions, 
\begin{equation}
I_\text{MF}(t) = \frac{1}{V_\text{DC}}\frac{dE_\text{MF}}{dt}. \label{Eq:Imf}
\end{equation}
See Appendix~B for derivation.
% We note that $E_\text{MF}(t)\propto\langle Q(t)|\sigma_z|Q(t)\rangle$ for a Majorana qubit, where subscripts are omitted. Since the rotation axis $\hat{r}$ of the Majorana qubit is not aligned along $\hat{z}$, the periods of Majorana qubit and $E_\text{MF}(t)$ are the same so that $I_\text{MF}$ measures the period of a Majorana qubit. 
%
%We elaborate on the relation between the Majorana qubit and $E_\text{MF}$. For a simple case when two Majorana fermions fuse and form a complex fermion state $|0\rangle$ or $|1\rangle$, we require a positive energy to excite $|1\rangle$ while we subtract an energy for its particle-hole partner $|0\rangle$. This opposite sign of excitation energies of opposite fermion parity states results in the opposite direction of supercurrent. Now with four Majorana fermions among which only two Majorana fermions fuse, the energy of complex fermions states $|00\rangle$ and $|11\rangle$ is determined by the fermion parity of two fused Majorana fermions. Although 
The energy $E_\text{MF}$ can be analytically derived, by using the effective Hamiltonian  $H_{int}(t) = i\sum_{j=1}^{3} E_j(t)\gamma_{j+1}(t)\gamma_{j}(t) + iE_4(t)\gamma_4(t)\gamma_1(t)$ of the four Majorana fermions, where the energy parameters $E_{j=1,2,3,4}$'s are evaluated from the overlap of the wave functions of the Majorana fermions. The effective Hamiltonian can be obtained~\cite{ChoiSim} from the Hamiltonian $H$, by projecting $H$ onto the space of the basis states.
%We provides the analytic expression of $nT_\text{J}$-periodic $I_\text{MF}(t)$ for $n=2,3,4$ using Eq.~\eqref{Eq:Imf} and the Hamiltonian involving interactions among the Majorana fermions $H_{int}(t) = i\sum_{j=1}^{3} s_j E_j(t)\gamma_{j-1}(t)\gamma_{j}(t) + iE_0(t)\gamma_3(t)\gamma_0(t)$. The energy splitting $E_j$ by the fusion of Majorana fermions $\gamma_{j-1}$ and $\gamma_{j}$ is evaluated from the overlap of wave functions of the Majorana fermions. 
For the initial state $|0_{41}0_{32}\rangle_0$ and at $\phi_+ = \pi$, $I_\textrm{MF}$ is obtained as
\begin{equation}\label{Eq:IMFn2}
I_\text{MF}(t) 
= 
\frac{eD_{\delta}}{\hbar}\delta
\sin\frac{\pi  t}{2 T_\text{J}} \sinh \left(\delta  \cos\frac{\pi  t}{2 T_\text{J}}\right), \,\, n=2.
\end{equation}
which is $2T_\text{J}$-periodic.
At  $\phi_+ = 5\pi/4$ and $\phi_+ = 3\pi/2$, $I_\text{MF}(t)$ is $3T_\textrm{J}$-periodic and $4T_\textrm{J}$-periodic, respectively, as 
\begin{eqnarray}
\!\!\!\!\!\!\!\! I_\text{MF}(t) \! = \!\! \left\{
{\setlength\arraycolsep{1pt}
\begin{array}{l}
\! \frac{e D_{9\delta/4}}{\hbar} \delta \sin \frac{\pi t}{3 T_\text{J}}\sinh \left(\frac{9\delta}{4}\! \cos \frac{\pi t}{3 T_\text{J}}\!\right)\!, \,\,\, n=3,\\
\\
\! \frac{e D_{\delta}}{\hbar} \delta \sin \frac{\pi t}{2 T_\text{J}}\cosh \left(\delta  \cos \frac{\pi t}{2 T_\text{J}}\right) ,\,\,\,\,\,\,\,\,\,  n=4.
\end{array}
}
\right. \label{Eq:IMFn3n4}
\end{eqnarray}
Here $D_\delta = \sqrt{ N(\hbar v_\textrm{J})^2 / [2l W I_0 (2 \delta)]}$ and $I_{\zeta=0}$ is the modified Bessel function of the first kind.

We expand Eq.~\eqref{Eq:IMFn2} and \eqref{Eq:IMFn3n4} into sinusoidal functions for later use. For the $n=2$ case, Eq.~\eqref{Eq:IMFn2} becomes
\begin{equation} \label{Eq:IMFn2-2}
I_\text{MF}(t)=\frac{2eD_\delta}{\hbar}
\sum_{p=1}^\infty 2pI_{\zeta=2p}(\delta)\sin\frac{p\omega_\text{J}t}{2}, \,\, n=2.
\end{equation}
where $\omega_\text{J}=2\pi/T_\text{J}$.
% and $I_{2p}(\delta)$ is the modified Bessel function of the first kind.
For the $n=3$ and $n=4$ cases, Eq.~\eqref{Eq:IMFn3n4} becomes
\begin{eqnarray} \label{Eq:IMFn3n4-2}
I_\text{MF}(t) = \left\{
{\setlength\arraycolsep{1pt}
\begin{array}{l}
\frac{2eD_{9\delta/4}}{\hbar} \sum\limits_{p=1}^\infty 2pI_{2p}\left(\frac{9\delta}{4}\right)\sin\frac{p\omega_\text{J}t}{3}, \quad n=3,\\
\\
\frac{2eD_{\delta}}{\hbar} \sum\limits_{p=1 \atop p \text{ odd}}^\infty p I_{p}\left(\delta\right)\sin\frac{p\omega_\text{J}t}{4}, \quad\qquad n=4.
\end{array}
}
\right.
\end{eqnarray} 
The frequency component with $p=1$ provides $nT_\text{J}$-periodic current, while the higher harmonics with $p\ge 2$ does $nT_\text{J}/p$-periodic current. Only several higher harmonics components contribute to $I_\text{MF}(t)$ since $I_{2p}\sim\exp(-2p^2/\delta)$. We note that the $4T_\text{J}$-periodic $I_\text{MF}(t)$ excludes the even harmonics, i.e., $p=1,3,5,\cdots$, since $I_\text{MF}(t+2T_\text{J})=-I_\text{MF}(t)$ when $n=4$ [see Eq.~\eqref{Eq:IMFn3n4}].
This feature is a generic feature of the $n=4$ case; in this case, 
$U^2 \propto \left( \begin{array}{cc} 0 & 1 \\ 1 & 0 \end{array} \right)$ and $U^4$ is proportional to the identity matrix.

%this feature does not rely on the specific functional form of $I_\text{MF}(t)$ or the initial state.
% Under $V_\text{DC}$ resulting in $n=4$, $U^2$ flips the fermion parity $|00\rangle\leftrightarrow|11\rangle$ so that $I_\text{MF}(t+2T_\text{J})=-I_\text{MF}(t)$ holds generally.

\section{Robustness of the $2n\pi$ fractional AC Josephson effects}\label{Sec:Practical}

As discussed in the above section~\ref{Subsec:FJ}, the non-Abelian state evolution of the Josephson junciton in Fig.~\ref{Fig:setup}
and the resulting $2n \pi$ fractional AC Josephson effects were found in Ref.~\cite{ChoiSim}, focusing on the states near the zero energy. 
In this section, we discuss that the $2n \pi$ fractional AC effects do not require fine tuning of the parameters
and that the $2n \pi$ effects can occur in the setup with realistic parameters, taking the effect of the midgap states of the junction into account. 

The $2n \pi$ fractional AC Josephson effect is insensitive to continuous deformation of the setup such as the shape of the arcs and different arc-lengths of the SCs. It is because the formation of the extended chiral Majorana zero modes along the arcs is of topological origin; the chiral modes occur along the boundary between the SC regions (which is topological) and the gapped TI region (which is topologically trivial). 

The $2n \pi$ effect does not require the fine tuning of the external magnetic field or the magnetic flux $N$ through the Josephson junction. In Ref.~\cite{ChoiSim}, it was proposed to apply $N=3$ magnetic flux quanta to the junction, to have the $2n \pi$ effect.
We find that the $2n \pi$ effect can still occur in the domain of  $N \in (2.6, 3.4)$.
The domain is estimated as follows. When the magnetic field becomes weaker so that the magnetic flux deviates from $N=3$ to $N = 3 - \delta N$, the separation between the localized Majorana zero modes increase. To host three localized Majorana zero modes inside the junction, the distance from the zero modes to the junction ends should be longer than the localization length of the zero modes.
On the other hand, when the magnetic field becomes stronger so that the magnetic flux deviates from $N=3$ to $N=3 + \delta N$, the separation between the localized Majorana zero modes decrease. In this case, the junction should not host four localized Majorana zero modes at any time.
By using Eq.~\eqref{Eq:positionMF}, the above conditions are fulfilled when $\delta N < \lambda N^2 / W \sim 0.4$ with $N=3$.
Note that $\delta N = 0.4$ corresponds to $\pm 17 \, \text{mT}$, if the magnetic field is $130 \, \text{mT}$ at $N=3$.

Below we will show in Sec.~\ref{Subsec:midgap} that the $2n \pi$ effects can occur in the setup with realistic parameters, taking the effect of the midgap states of the junction into account.
We will also show in Sec.~\ref{Subsec:Vdc} that the $2n \pi$ fractional AC effects do not require fine tuning of the DC voltage $V_\textrm{DC}$ across the Josephson junction.

%Ref.~\cite{ChoiSim} has studied various Majorana qubit rotations, invoking the non-Abelian fusion of four Majorana fermions with a DC voltage bias $V_\text{DC}$, and suggested experimental detection schemes utilizing the anomalous Josephson current. In this section, we take into account of realistic considerations for experimental feasibility in the following subsections. In Sec.~\ref{Subsec:midgap}, we show that with realistic experimental parameters, the anomalous Josephson current can be comparable to the conventional Josephson currents so that the  $2n\pi$ fractional Josephson effect is measurable experimentally. We focus on $N=3$ without vortices ($M_\text{u}=M_\text{l}=0$) where nonzero $I_\text{MF}$ exists. In Sec.~\ref{Subsec:Vdc}, we provide that in achieving a desired Majorana qubit rotation, it is not crucial to attain a fine tuning of $V_\text{DC}$. 

\subsection{Contribution from the midgap states}\label{Subsec:midgap}

The Josephson junction has the midgap states with nonzero energy smaller than the superconducting gap $\Delta_0$ in addition to the Majorana zero modes. These midgap states are topologically trivial and result in the $2 \pi$ periodic contribution to the Josephson current.
To observe the $2n\pi$ fractional AC Josephson effects, the $2 n \pi$ periodic contribution from the Majorana zero modes has to be at least comparable to the $2 \pi$ periodic contribution from the midgap states.
In this section, we will show that there is a realistic parameter regime in which the $2n \pi$ contribution is sufficiently large, in comparison with the $2 \pi$ contribution, so that the $2 n \pi$ fractional AC Josephson effect can be observed.

We begin with the estimation of the $2n \pi$ contribution $I^{max}_\text{MF}$ by the Majorana zero modes
to the Josephson current. The amplitude of the $2n \pi$ contribution is approximately  $I^{max}_\text{MF}\approx\frac{2e}{\hbar}\frac{\delta E_\text{MF}}{\delta\varphi_0}$, where $\delta E_\text{MF}$ is the energy increment by
the fusion or splitting of the Majorana zero modes discussed in Sec.~\ref{Subsec:FJ}.
The fusion and splitting occur, at the junction ends, between a localized Majorana zero mode inside the junction and the extended Majorana zero mode along the arcs outside the junction. The resulting maximum energy gain or loss is estimated as $\delta E_\text{MF}\approx \hbar v_\text{arc}\pi/(l_\text{u}+l_\text{l})$ which is essentially the half of the level spacing of the extended chiral modes along the arcs.
The energy gain or loss occurs while the superconducting phase difference $\varphi_0$ changes by $\delta \varphi_0\approx N\pi\lambda_M/W$ with which the localized Majorana zero mode moves to a junction end by its localization length $\lambda_M$
[see Eq.~\eqref{Eq:positionMF}].
%  The maximum energy splitting is $\delta E_\text{MF}\approx v_\text{arc}\pi/(l_\text{u}+l_\text{l})$, since it is produced as the localized Majorana fermion forms the complex fermion state encircling the arcs of SCs at the finite energy according to Eq.~\eqref{Eq:QuantRule}.  The phase evolution during the fusion is obtained as $\delta \varphi_0\approx N\pi\lambda_M/W$ from Eq.~\eqref{Eq:positionMF} by noting that the fusion occurs as a localized Majorana fermion moves from $x_k=W-\lambda_M$ to $x_k=W$. 
Consequently, the amplitude of the $2n \pi$ contribution is estimated as
\begin{equation}
I_\text{MF}^{max} \approx  
\frac{2e v_\text{arc}W}{N(l_\text{u}+l_\text{l})\lambda_M}.
\end{equation}
In a circular geometry, $\pi W = (l_\textrm{u} + l_\textrm{l})$. And $v_\text{arc}\sim v_F$ under typical parameters.~\cite{Bruder} Then,
 %To increase $I_\text{MF}^{max}$, sharply localized Majorana fermions ($\lambda_M < W$) is required to fuse into the complex fermion states with large energy level spacing ($v_\text{arc} > l_\text{u}+l_\text{l}$).  With typical parameters, $v_\text{arc}\sim v_F$. We note that the arcs of SCs need to be spatially separated from the junction to avoid unnecessary hybridization among chiral modes $\eta_a$, which corresponds to $l_\text{u}+l_\text{l}>W>\xi=v_F/\Delta_0$, where $\xi$ is the superconducting coherence length. Assuming the circular geometry in Fig.~\ref{Fig:setup}, we obtain
\begin{equation}
I_\text{MF}^{max}\approx\frac{2e\Delta_0}{\hbar}\frac{\xi}{N\pi\lambda_M}, \label{Eq:ImfMax}
\end{equation}
where $\xi = \hbar v_F / \Delta_0$ is used.

Next, we estimate the amplitude $I_\text{mid}^{max}$ of the $2 \pi$ contribution from the midgap states of nonzero energy.
Inside the junction, such midgap states occur at the same positon with the localized Majorana zero modes.
They have a longer localization length $\lambda_\nu \sim \sqrt{\nu} \lambda_M$ than the zero modes, where $\nu = 1,2, \cdots$. 
As the SC phase difference $\varphi_0$ changes, the midgap states move. When they move to the junction end and merge with the extended midgap states propagating along the arcs, their energy changes and contributes to the Josephson current.
Similarly to the case of the Majorana zero modes,
the energy change is estimated as $\delta E_\nu \approx 2 \hbar v_\text{arc}\pi/(l_\text{u}+l_\text{l})$, 
and it occurs when the SC phase difference changes by  $\delta\varphi_\nu \approx N\pi \lambda_\nu /W = N\pi\sqrt{\nu}\lambda_M/W$.
The level spacing of the merged midgap states is approximately $\delta E_\nu$, hence, the number of those states within the SC gap is $\nu_{max} \simeq \Delta_0 / \delta E_\nu$.
Combining the above facts, the amplitude $I_\text{mid}^{max} = (2e / \hbar) \sum_{\nu=1}^{\nu_{max}}\delta E_\nu / \delta\varphi_\nu$ of the $2 \pi$ contribution from the midgap states is estimated as
% These Andreev bound states gives rise to $I_\text{mid}$ when they move from $x_k=W-\lambda_\nu$ to $x_k=W$ becoming the complex fermion state along the arcs of SCs at the finite energy $E_\nu=(2\nu+1)\pi v_\text{arc}/(l_\text{u}+l_\text{l})$. Similarly to the estimation of $I_\text{MF}^{max}$ using Eq.~\eqref{Eq:positionMF} and \eqref{Eq:QuantRule}, 
\begin{eqnarray}
I_\text{mid}^{max} & \approx & \frac{2e\Delta_0}{\hbar}\frac{\xi}{N\pi\lambda_M} \sum_{\nu=1}^{\nu_{max}}\frac{1}{\sqrt{\nu}} \nonumber \\
& < &  \frac{2e\Delta_0}{\hbar}\frac{\xi}{N\pi\lambda_M} \times 2.8  \left(\sqrt{\frac{W}{\xi}}-1\right), \label{Eq:IabsMax}
\end{eqnarray}
%where $\nu_{max} \approx (\Delta_0 W/v_\text{arc}-1)/2$. 
where $\pi W = (l_\textrm{u} + l_\textrm{l})$ was used, considering the case of the circular geometry, and a partial sum of a Riemann zeta function was used to get the upper bound of the estimation.

%%%%%%%%%%%%%%%%%%%%%%%%%%%%%%%%%%%%% FIGURE BEGIN %%%%%%%%%%%%%%%%%%%%%%%%%%%%%%%%%%%%%

\begin{figure}[pt]
\centering
\includegraphics[width=0.99\columnwidth]{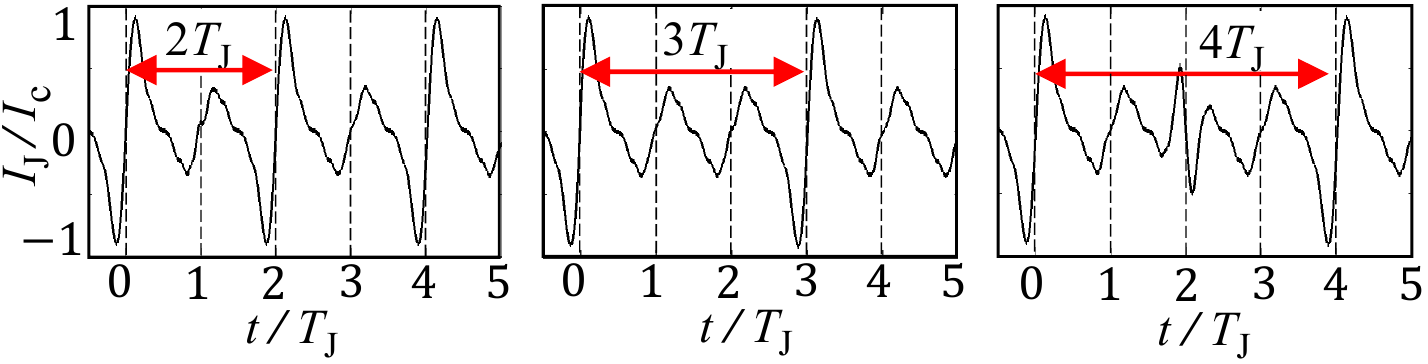}
\caption{Josephson current $I_\text{J}$ (the sum of the contribution from the Majorana zero modes and that from the other midgap states) as a function of time $t$ for $n=2,3,4$. 
%The parameters suggesed in Ref.~\cite{ChoiSim} are used:
$\phi_+$ and the corresponding value of $V_\textrm{DC}$ are chosen as
$\phi_+ = \pi$ and $V_\textrm{DC} = 42 \, \mu$V in the left panel for $n=2$,
$\phi_+ = 3\pi/4$ and $V_\textrm{DC} = 56 \, \mu$V in the middle panel for $n=3$, and
$\phi_+ = \pi/2$ and $V_\textrm{DC} = 84 \, \mu$V  in the right panel for $n=4$. 
%The relation between $\phi_+$ and $V_\textrm{DC}$ is $\phi_+ = \pi  \overline{E}/(e V_\textrm{DC})$, and $  \overline{E} = 42 \, \mu$V is obtained for the parameters suggested in Fig.~\cite{ChoiSim}. 
$|0_{41} 0_{32}\rangle_0$ is chosen as the initial state at $t=0$.
The other parameters are chosen as in Fig.~\ref{Fig:MFsN3}.
The Josephson current exhibits $2n\pi$ fractional AC Josephson effects.
 }\label{Fig:midgap}
\end{figure}
%%%%%%%%%%%%%%%%%%%%%%%%%%%%%%%%%%%%% FIGURE END %%%%%%%%%%%%%%%%%%%%%%%%%%%%%%%%%%%%%
%Moreover, we numerically find $I_\text{mid}^{max} \sim I_\text{MF}^{max}$ by computing each contribution of $I_\text{mid}^{max}$ and $I_\text{MF}^{max}$ with $H$ in Eq.~\eqref{Eq:Heff} and parameters suggested in Fig.~\ref{Fig:MFsN1N2}, showing the $2n \pi$ fractional Josephson effect. The numerical results of the total Josephson current $I_\text{J}=I_\text{MF}+I_\text{mid}$ for $n=2,3,4$ are shown in Fig.~\ref{Fig:midgap}. 
%We obtain the Majorana qubit rotation by numerically evaluating the time evolution operator $U(t,0)$. We use $U(t,0)=U(t,t-\Delta t)U(t-\Delta t,t-2\Delta t)\cdots U(\Delta t,0)$ and $U(k\Delta t+\Delta t,k\Delta t) \simeq \exp( -\frac{i\Delta t}{\hbar}H_\text{lattice}(k\Delta t) )$, where $\Delta t = t/N_\text{time}$ and $k=0,\cdots, N_\text{time}-1$. The total energy is obtained by $E_\text{tot}(t) = \sum_k E_k(t)|\langle k(t) | U(t,0) | \psi(t=0) \rangle|^2$, where $E_k(t)$ and $|k(t)\rangle$ satisfy $H_\text{lattice}(t) |k(t)\rangle = E_k(t)|k(t)\rangle$, and $k$ is the level index of the states of the junction (including those from the four Majorana fermions and those from the midgap states in the bulk SC gap). The total Josephson current is evaluated by $I_J(t) = \frac{1}{V_\text{DC}}\frac{\partial E_\text{tot}}{\partial t}$.

We compare the $2n\pi$ contribution $I_\text{MF}^{max}$ and the upper bound of the $2 \pi$ contribution $I_\text{mid}^{max}$ in Eqs.~\eqref{Eq:ImfMax} and  \eqref{Eq:IabsMax} in the parameter regime of $\xi \le W / N$ and $N=3$.
This regime of $\xi \le W / N$ is chosen, because in this regime the localized Majorana zero modes form nonlocal fermion states robust against unwanted local processes such as parity flip.
When $W = 3 \xi$, the $2n \pi$ contribution is sufficiently large, $I_\text{MF}^{max} \sim 0.5 I_\text{mid}^{max}$, in comparison with the $2\pi$ contribution.
We numerically compute the total Josephson current in Fig.~\ref{Fig:midgap}, which shows that
 the $2n \pi$ contribution is comparable to the $2\pi$ contribution, $I_\text{MF}^{max} \simeq I_\text{mid}^{max}$.
This confirms that there is the realistic parameter regime in which the $2n \pi$ contribution is sufficiently large so that the $2 n \pi$ fractional AC Josephson effect can be observed.

We note that the numerical computation is done, solving the time-dependent Schrodinger equation with
the effective Hamiltonian in Eq.~\eqref{Eq:Heff}

%Note that the contribution of $I_\text{mid}$ increases when the system size becomes larger and accommodates more midgap states. Although $I_\text{mid}^{max}$ vanishes in Eq.~\eqref{Eq:IabsMax} when $W=\xi$, this situation ruins the conservation of local fermion parity of multiple localized Majorana fermions (e.g. $N=3$). 
%When the distance between localized Majorana fermions $W/N$ is smaller than the size of cooper pair $\xi$, a pair of local fermion parity can flip as tunneling into a cooper pair without non-Abelian effect; $|00\rangle\leftrightarrow|11\rangle$.
% Using $\xi \le W/N$ to avoid the unfavorable process, we estimate $I_\text{mid}^{max}\le 2I_\text{MF}^{max}$ with $N=3$. 

%In realistic situations, the conventional Andreev bound states in the junction appear. Their contribution to the Josephson current $I_\text{mid}$ is $2 \pi$ periodic in contrast to the anomalous Josephson current $I_\text{MF}$ from the four Majorana fermions which can be $2 n \pi$ periodic. Provided that the maximum of $I_\text{MF}$ is not much smaller than that of the midgap states, the total Josephson current, the sum of the two contributions, is $2n \pi$ periodic, showing the $2n \pi$ fractional Josephson effects. In this subsection, we estimate the maximal value of $I_\text{MF}$ and $I_\text{mid}$ as $I^{max}_\text{MF}$ and $I^{max}_\text{ABS}$, respectively, to show under which condition we can obtain $I_\text{MF}$ comparable to $I_\text{mid}$.

\subsection{$V_\textrm{DC}$ dependence of the $2n\pi$ AC Josephson effects}\label{Subsec:Vdc}
%We discuss the dependence of $I_\text{J}(t)$ on $V_\text{DC}$.
%As shown in the subsection~\ref{Subsec:Qubit}, $\phi_+=\pi\overline{E}/(eV_\text{DC})$ determines the period of AC Josephson current, while  $\overline{E}$ is independent of  $V_\text{DC}$ but depends on the geometry of the setup so that $V_\text{DC}$ can be the tuning knob of the period. 

In Ref.~\cite{ChoiSim}, it was shown that the period-elongation factor $n$ of the $2n \pi$ fractional AC Josephson effects can be tuned by changing the DC voltage $V_\textrm{DC}$ across the Josephson junction. 
In this section, we show that the $2n \pi$ effects of a given value of $n$ can occur in a range of $V_\textrm{DC}$ (not only at a specfic value of $V_\textrm{DC}$). Therefore the $2n \pi$ fractional AC Josephson effects do not require fine tuning of the DC voltage $V_\text{DC}$ across the junction.

%The DC voltage bias across the junction $V_\text{DC}$ is the experimental tuning knob which determines the period of  the fraction $2n\pi$ AC Josephson current.

%We show the dependence of $I_\text{J}(t)$ as changing $V_\text{DC}$ continuously. We find that the possible experimental deviation of $V_\text{DC}$ does not change $I_\text{J}(t)$ abruptly within a finite period of time so that the $2n\pi$ fractional AC Josephson effect sustains.

%%%%%%%%%%%%%%%%%%%%%%%%%%%%%%%%%%%%% FIGURE.2 BEGIN %%%%%%%%%%%%%%%%%%%%%%%%%%%%%%%%%%%%%
\begin{figure}[b]
\centering
\includegraphics[width=0.99\columnwidth]{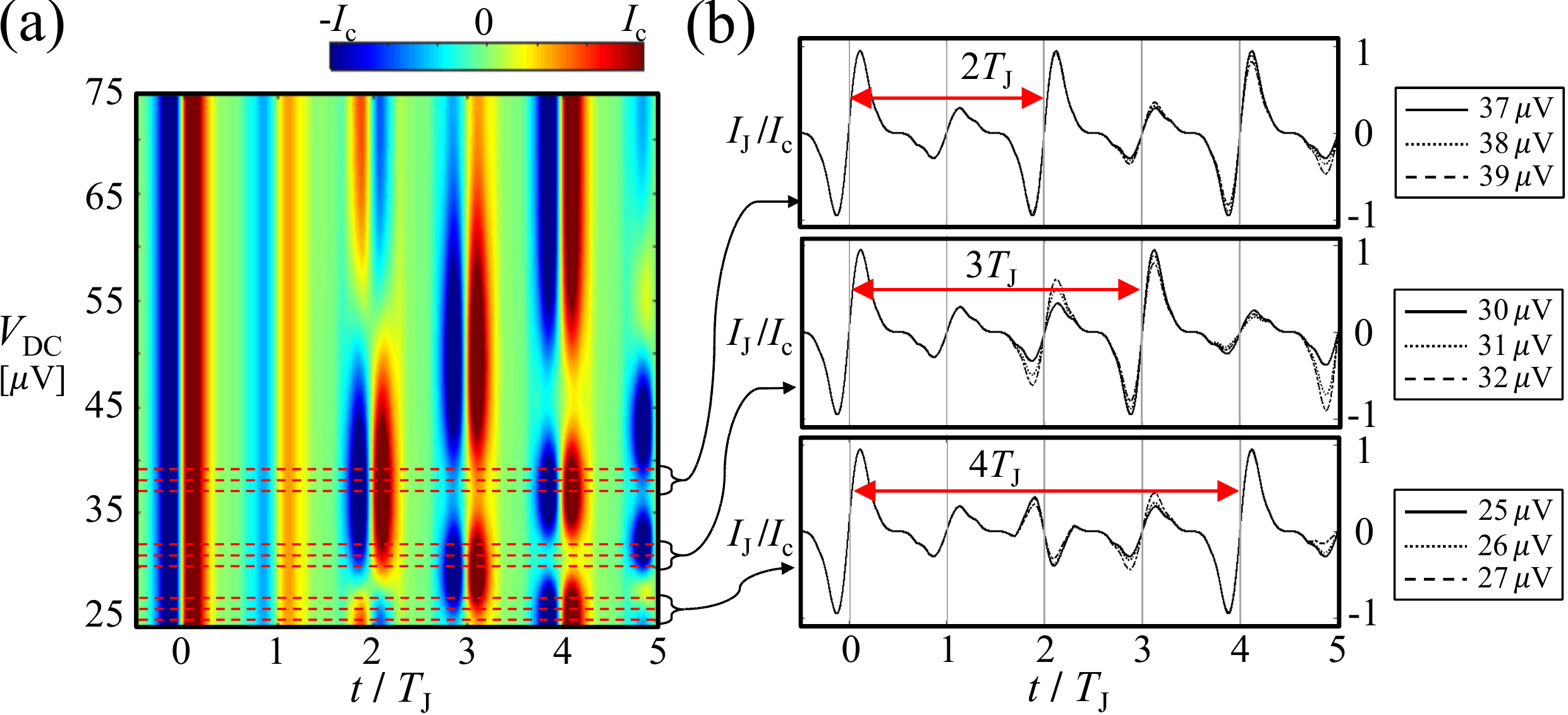}
\caption{(a) Colormap of the Josephson current $I_\text{J}(t)$ as a function of $V_\text{DC}$, numerically computed with  the parameters suggested in the Fig.~\ref{Fig:MFsN3} and the initial state $|0_{41}0_{32}\rangle_0$ at $t=0$. 
The contribution from the midgap states is included in $I_\text{J}(t)$. 
Guiding red dashed lines are drawn at $V_\text{DC}=25, 26, 27, 30, 31, 32, 37, 38, 39 \, \mu\text{V}$. For these $V_\text{DC}$'s, $I_\text{J}(t)$ is presented in (b). 
The period of $I_\text{J}(t)$ is $2T_\text{J}$, $3T_\text{J}$, and $4T_\text{J}$ in the top, middle, and bottom panels, respectively. 
This shows that  the $2n \pi$ effects of a given value of $n$ can occur in a range of $V_\textrm{DC}$ (not only at a specfic value of $V_\textrm{DC}$).
} \label{Fig:Vdc}
\end{figure}
%%%%%%%%%%%%%%%%%%%%%%%%%%%%%%%%%%%%% FIGURE.2 END %%%%%%%%%%%%%%%%%%%%%%%%%%%%%%%%%%%%%

In Fig.~\ref{Fig:Vdc}(a), we numerically compute the Josephson current $I_\text{J}(t)$ for different values of $V_\text{DC}$, including the contribution from the midgap states.
The result shows that $I_\text{J}(t)$ smoothly changes with $V_\text{DC}$.
For example,  at the different values of $V_\text{DC}=37, 38, 39 \, \mu\text{V}$, the period of $I_\text{J}(t)$ can be considered as $2T_\text{J}$ within a range of time.
It is because the evolution of  the four Majorana fermions $\gamma_k$'s (or the rotation axis $\hat{r}$ and rotation angle $\Omega$) changes only slightly in a range $\Delta V_\textrm{DC}$ of $V_\textrm{DC}$ ($\Delta V_\textrm{DC}$ will be estimated below).
Similarly, the period of $I_\text{J}(t)$ is $3 T_\text{J}$ around $V_\textrm{DC} = 30 \, \mu \text{V}$
and it is $4 T_\text{J}$ around $V_\textrm{DC} = 25 \, \mu \text{V}$.

%For example,  the period of $I_\text{J}(t)$ is $2T_\text{J}$ at $V_\text{DC}=37 \, \mu\text{V}$ where $V_\text{DC}=\overline{E}/e$ and equivalently $\phi_+ = \pi$. In this case, the qubit formed by the four Majorana fermions $\gamma_k$'s rotates about the axis $\hat{r}$ by the angle $\Omega = \pi$ during one $T_\text{J}$, and it recovers the initial state at $t=2 T_\text{J}$ after rotating by $2 \Omega = 2 \pi$. When $V_\text{DC}$ deviates slightly from $37 \, \mu\text{V}$, for example, to $38 \, \mu\text{V}$ [see Fig.~\ref{Fig:Vdc}(b)], $\hat{r}$ and $\Omega$ changes also slightly. As a result, the current  $I_\text{J}(t)$ at $38 \, \mu\text{V}$ differs slightly from that of $37 \, \mu\text{V}$ over multiples of $T_\text{J}$'s.
 If we evaluate the period with an unrealistic perfect accuracy of $I_\text{J}(t)$ in the case of  $V_\text{DC}=37, 38, 39 \, \mu\text{V}$, the exact period of $I_\text{J}(t)$  can largely deviate from $2 T_\text{J}$. While the exact period is $2 T_\text{J}$ at $37 \, \mu\text{V}$, the period is $27T_\text{J}$ at $38 \, \mu\text{V}$ and $69T_\text{J}$ at $39 \, \mu\text{V}$ in Fig.~\ref{Fig:Vdc}(b).  Similar phenomena happen around $V_\text{DC}=30 \, \mu\text{V}$ and $V_\text{DC}=25 \, \mu\text{V}$; note that the exact periods of $3T_\text{J}$ and $4T_\text{J}$ occur at  $V_\text{DC}=30 \, \mu\text{V}$ and $V_\text{DC}=25 \, \mu\text{V}$, respectively. Accordingly, we below show that the periodicity $2n\pi$ of the current can be defined with a small fraction of tolerance in $I_\text{J}(t)$.

As shown that $I_\text{J}(t)$ changes smoothly with $V_\text{DC}$, we quantify the smoothness and relate it to the observability of the $2n\pi$-periodic fractional AC Josephson effect.
We consider the voltage $V_\textrm{DC}^{(n)}$ at which the Josephson current is exactly $n T_\textrm{J}$ periodic in time, and the current is denoted by $I^{(n)}_\textrm{J}(t)$. When the voltage is $V_\textrm{DC} = V_\textrm{DC}^{(n)} + \Delta V_\textrm{DC}$, the time dependence of the current changes to $I_\text{J}(t)$. We quantify the difference between $I^{(n)}_\textrm{J}(t)$ and $I_\textrm{J}(t)$ by
\begin{equation}
\epsilon(t) = 1-\left|\frac{\int_0^t dt' I_\text{J}(t') I^{(n)}_\text{J}(t')}{\int_0^t dt' [I^{(n)}_\text{J}(t')]^2}\right|.
\end{equation}
We consider time $t_{10 \%}$ at which $\epsilon(t_{10 \%}) = 0.1$ (within which we have $10 \%$ tolerance). 
%Then, we can define the maximum period of time $t_\text{max}$ during which approximately $nT_\text{J}$ period can be measured within $10\%$ tolerance; $t_\text{max} \equiv \max \{ t | \epsilon(t) < 10\% \}$. 
We estimate it, %{\color{blue} upto the first order of $V^{(n)}_\text{DC}/|\Delta V_\text{DC}|$}, 
\begin{equation}
t_{10 \%}  = \frac{\pi }{2 \phi_+^{(n)}}\left(1+\frac{V_\textrm{DC}^{(n)}}{|\Delta V_\text{DC}|}\right) T_\text{J} + \mathcal{O} (\frac{{V_\textrm{DC}^{(n)}}^2}{|\Delta V_\text{DC}|^2}).
\end{equation}
%$t_\text{max}/T_\text{J} \sim \frac{eV^0_\text{DC}}{2\overline{E}}\left(1+\frac{V^0_\text{DC}}{\Delta V_\text{DC}}\right)$, 
$\phi_+^{(n)}$ is the value of $\phi_+$ at $V_\textrm{DC}^{(n)}$.
For example, when the period is $2 T_\textrm{J}$, $\phi_+^{(n=2)}= \pi$ and  $t_{10 \%} \sim 5 T_\text{J}$ at $|\Delta V_\text{DC}| = 0.1 V_\textrm{DC}^{(n=2)}$, implying that the current $I_\text{J}(t)$ is almost identical to $I_\text{J}^{(2)}(t)$ and hence can be considered as $2 T_\textrm{J}$ periodic within the time intervals of $5 T_\textrm{J}$. 
This shows that a $4 \pi$ fractional AC Josephson effect can be observed in the voltage interval of 33 - 41 $\mu\text{V}$ around $V_\textrm{DC}^{(2)} = 37 \, \mu\text{V}$ in the parameter regime of Fig.~\ref{Fig:Vdc}.
Similarly, when the period is $3 T_\textrm{J}$, $\phi_+^{(3)}= 5\pi/4$ and $t_{10 \%} \sim 5 T_\text{J}$ at $|\Delta V_\text{DC}| = 0.09 V_\textrm{DC}^{(3)}$,  which shows that a $6 \pi$ fractional AC Josephson effect can be observed in the interval of 27 - 33 $\mu\text{V}$ around $V_\textrm{DC}^{(3)} = 30 \, \mu\text{V}$.
When the period is $4 T_\textrm{J}$, $\phi_+^{(4)}= 3\pi/2$ and $t_{10 \%} \sim 5 T_\text{J}$ at $|\Delta V_\text{DC}| = 0.07 V_\textrm{DC}^{(4)}$,  which shows that a $8 \pi$ fractional AC Josephson effect can be observed in the interval of 23.2 - 26.8 $\mu\text{V}$ around $V_\textrm{DC}^{(4)} = 25 \, \mu\text{V}$.

This demonstrates the possibility of observing our $2n\pi$-periodic fractional AC Josephson effect of at least small $n \ge 2$ in sufficiently wide windows of $V_\text{DC}$, hence, the possibility of observing that $n$ changes with $V_\text{DC}$. This dependence of $n$ on $V_\text{DC}$ allows us to distinguish our $2n\pi$ fractional Josephson effect
from the other fractional AC Josephson effects in which $n$ is independent of $V_\text{DC}$.

\section{Non-commutativity of Majorana zero modes in the Josephson junction}\label{Sec:NonCom}
%In Ref~\cite{ChoiSim}, it has been shown that the $2n\pi$ fractional AC Josephson effect appears because of the non-Abelian evolution of a qubit state formed by the four Majorana fermions. The non-Abelian evolution is described by the operator $U$. 
The $2n\pi$ fractional AC Josephson effect offers the simplest way of observing the non-Abelian evolution of the Josephson junction.
Below, we provide another approach based on the Josephson junction,
%This approach is more complicated than the measurement of the $2n\pi$ fractional AC Josephson effect, but it allows us to observe the non-commutativity of the operation $U$, hence, 
which offers a more direct way of observing the non-Abelian nature of the Majorana zero mode.  

\begin{figure*}[t]
\centering
\includegraphics[width=1.4\columnwidth]{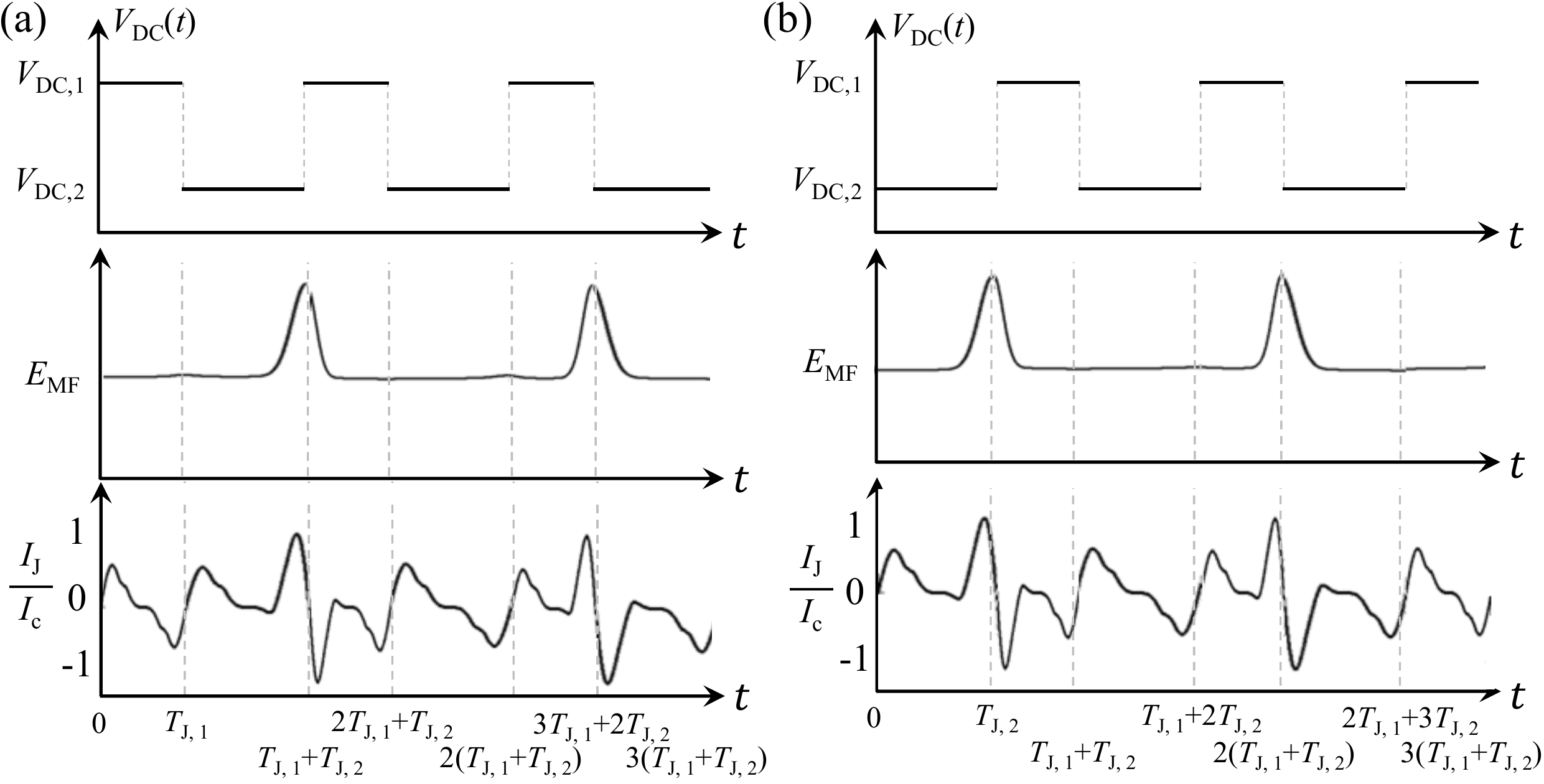}
\caption{An approach of observing the non-commutativity of the operation $U$. The case ``21'' in (a) and the case `'12'' (b) are compared. For each case, the top panel shows the time dependence of $V_\text{DC}$ (time intervals of $V_{\textrm{DC},1}$ and those of $V_{\textrm{DC},2}$ are alternately applied as indicated by vertical dashed lines), and the second and third panels show the numerical calculation results of the energy $E_\text{MF}(t)$ of the state formed by the four Majorana fermions $\gamma_{k=1,2,3,4}$ and  the total Josephson current $I_\text{J}(t)$ (which includes the contribution from the midgap states).
We use $V_{\textrm{DC},1}=37 \, \mu\text{V}$, $V_{\textrm{DC},2}=25 \, \mu\text{V}$, and the other parameters suggested in Fig.~\ref{Fig:MFsN3}. The same initial state $\frac{1}{\sqrt{2}}\left(|0_{41}0_{32}\rangle_0+|1_{41}1_{32}\rangle_0\right)$ is chosen at $t=0$ in both of (a) and (b).
} \label{Fig:V1V2}
\end{figure*}

In this approach, the voltage $V_{\textrm{DC}}$ is time dependent. Two constant values of the voltage, $V_{\textrm{DC},1}$ and $V_{\textrm{DC},2}$, are applied alternately in two different ways ``21'' and ``12'' (see top panels in Fig,~\ref{Fig:V1V2}).
In ``21'', we apply $V_{\textrm{DC},1}$ during the first time interval %$t \in [0, T_{\textrm{J},1}]$ 
of length $T_{\textrm{J},1} = h / (2e V_{\textrm{DC},1})$, 
$V_{\textrm{DC},2}$ during the second interval %$[T_{\textrm{J},1}, T_{\textrm{J},1} + T_{\textrm{J},2}]$ 
of length $T_{\textrm{J},2} = h / (2e V_{\textrm{DC},2})$, 
$V_{\textrm{DC},1}$ during the third interval 
%$[T_{\textrm{J},1} + T_{\textrm{J},2}, 2T_{\textrm{J},1} + T_{\textrm{J},2}]$ 
of length $T_{\textrm{J},1}$, 
$V_{\textrm{DC},2}$ during the fourth interval 
%$[2T_{\textrm{J},1} + T_{\textrm{J},2}, 2T_{\textrm{J},1} + 2T_{\textrm{J},2}]$ 
of length $T_{\textrm{J},2}$, and so on.
In ``12'', we apply $V_{\textrm{DC},2}$ during the first interval 
%$t \in [0, T_{\textrm{J},2}]$ 
of length $T_{\textrm{J},2}$, $V_{\textrm{DC},1}$ during the second interval 
%$[T_{\textrm{J},2}, T_{\textrm{J},1} + T_{\textrm{J},2}]$ 
of length $T_{\textrm{J},1}$, $V_{\textrm{DC},2}$ during the third interval
% $[T_{\textrm{J},1} + T_{\textrm{J},2}, T_{\textrm{J},1} + 2T_{\textrm{J},2}]$ 
of length $T_{\textrm{J},2}$, $V_{\textrm{DC},1}$ during the fourth interval 
%$[T_{\textrm{J},1} + 2T_{\textrm{J},2}, 2T_{\textrm{J},1} + 2T_{\textrm{J},2}]$ 
of length $T_{\textrm{J},1}$, and so on.

In the case ``21'', an initial state $|\psi \rangle$ of the four Majorana fermions $\gamma_{k=1,2,3,4}$ at time $t=0$ evolves into $U_2 U_1 |\psi \rangle$ at time $t=T_{\textrm{J},1} + T_{\textrm{J},2}$, and into  $(U_2 U_1)^m |\psi \rangle$ at $t=m(T_{\textrm{J},1} + T_{\textrm{J},2})$, where $m$ is a positive integer and $U_{i=1,2}$ is the evolution operator $U$ for each time interval of length $T_{\textrm{J},i}$ where the voltage $V_{\textrm{DC},i}$ is applied.
By contrast, in the case ``12'', the initial state evolves into $U_1 U_2 |\psi \rangle$ at time $t=T_{\textrm{J},1} + T_{\textrm{J},2}$, and into  $(U_1 U_2)^m |\psi \rangle$ at $t=m(T_{\textrm{J},1} + T_{\textrm{J},2})$.
The noncommutativity $U_2 U_1 \ne U_1 U_2$ can be observed by comparing the Josephson current between the cases ``21'' and ``12''.

%%%%%%%%%%%%%%%%%%%%%%%%%%%%%%%%%%%%% FIGURE.3 BEGIN %%%%%%%%%%%%%%%%%%%%%%%%%%%%%%%%%%%%%
%%%%%%%%%%%%%%%%%%%%%%%%%%%%%%%%%%%%% FIGURE.3 END %%%%%%%%%%%%%%%%%%%%%%%%%%%%%%%%%%%%%

This strategy is numerically demonstrated in Fig.~\ref{Fig:V1V2}. In the figure, we compare the time dependence of the energy of the state $|\psi (t) \rangle$ formed by the Majorana fermions and the Josephson current between the two cases ``21'' and ``12''. In both the cases, the evolution of the state starts from the same initial state $\frac{1}{\sqrt{2}}\left(|0_{41}0_{32}\rangle_0+|1_{41}1_{32}\rangle_0\right)$ at time $t=0$. In the case ``21'', the state evolves as 

%\begin{widetext}
%\begin{alignat*}{3}
%\frac{1}{\sqrt{2}}\left(|0_{41}0_{32}\rangle_0+|1_{41}1_{32}\rangle_0\right)
%\overset{U_2 U_1}{\longmapsto}   |1_{41}1_{32}\rangle_0 
%\overset{U_2 U_1}{\longmapsto}  \frac{1}{\sqrt{2}}\left(|0_{41}0_{32}\rangle_0+i|1_{41}1_{32}\rangle_0\right)
%\overset{U_2 U_1}{\longmapsto}  \frac{1}{\sqrt{2}}\left(|0_{41}0_{32}\rangle_0+|1_{41}1_{32}\rangle_0\right)\\
%\frac{1}{\sqrt{2}}\left(|0_{41}0_{32}\rangle_0+|1_{41}1_{32}\rangle_0\right)
%\overset{U_1 U_2}{\longmapsto} \frac{1}{\sqrt{2}}\left(|0_{41}0_{32}\rangle_0-i|1_{41}1_{32}\rangle_0\right)
%\overset{U_1 U_2}{\longmapsto} |1_{41}1_{32}\rangle_0
%\overset{U_1 U_2}{\longmapsto} \frac{1}{\sqrt{2}}\left(|0_{41}0_{32}\rangle_0+|1_{41}1_{32}\rangle_0\right)
%\end{alignat*}
%\end{widetext}

\begin{widetext}
\begin{alignat*}{3}
&\frac{1}{\sqrt{2}}\left(|0_{41}0_{32}\rangle_0+|1_{41}1_{32}\rangle_0\right)
&& \overset{U_2 U_1}{\longmapsto}   |1_{41}1_{32}\rangle_0 && \quad \textrm{at}   \quad t = T_{\textrm{J},1} + T_{\textrm{J},2},  \\
& && \overset{U_2 U_1}{\longmapsto}  \frac{1}{\sqrt{2}}\left(|0_{41}0_{32}\rangle_0+i|1_{41}1_{32}\rangle_0\right)
&& \quad \textrm{at} \quad  t = 2(T_{\textrm{J},1} + T_{\textrm{J},2}),  \\
& && \overset{U_2 U_1}{\longmapsto}  \frac{1}{\sqrt{2}}\left(|0_{41}0_{32}\rangle_0+|1_{41}1_{32}\rangle_0\right) && \quad \textrm{at} \quad  t = 3(T_{\textrm{J},1} + T_{\textrm{J},2}).
\end{alignat*}
In the case ``12'', the state evolves as 
\begin{alignat*}{3}
& \frac{1}{\sqrt{2}}\left(|0_{41}0_{32}\rangle_0+|1_{41}1_{32}\rangle_0\right)
&& \overset{U_1 U_2}{\longmapsto} \frac{1}{\sqrt{2}}\left(|0_{41}0_{32}\rangle_0-i|1_{41}1_{32}\rangle_0\right) && \quad \textrm{at} \quad t = T_{\textrm{J},1} + T_{\textrm{J},2}, \\
& && \overset{U_1 U_2}{\longmapsto} |1_{41}1_{32}\rangle_0 &&
 \quad \textrm{at} \quad  t = 2(T_{\textrm{J},1} + T_{\textrm{J},2}),  \\
& && \overset{U_1 U_2}{\longmapsto} \frac{1}{\sqrt{2}}\left(|0_{41}0_{32}\rangle_0+|1_{41}1_{32}\rangle_0\right) && \quad \textrm{at} \quad  t = 3(T_{\textrm{J},1} + T_{\textrm{J},2}).
\end{alignat*}
\end{widetext}

As a result, the energy and the Josephson current are different between the two cases ``21'' and ``12'' around $t = (T_{\textrm{J},1} + T_{\textrm{J},2})$. 
This offers a direct demonstration of the noncommutativity of $U_2 U_1 \ne U_1 U_2$.

\section{Shapiro spikes at large AC voltage bias %$eV_\text{AC}\sim\hbar\omega_\text{AC}$
}\label{Sec:Shapiro}

In Ref.~\cite{ChoiSim}, it was shown that the $2n \pi$ fractional AC Josephson effects can be detected by measuring Shapiro spikes with applying an AC voltage 
\begin{equation}
V(t) = V_\textrm{DC} + V_\textrm{AC} \sin (\omega t)
\end{equation}
of frequency $\omega$ across the junction, focusing on the weak AC voltage regime of  $eV_\text{AC}\ll \textrm{min} \{ 2e V_\textrm{DC}, \hbar\omega \}$. In this regime, the state evolution of the junction is affected only weakly by $V_\textrm{AC}$, and the main Shapiro spikes appear
when $2e V_\textrm{DC} / (\hbar \omega) = n / p$ is satisfied with certain integers $p \ge 1$; other possible spikes (which we call sub-spikes) at $2e V_\textrm{DC} / (\hbar \omega) = nq / p$ with $q = 2, 3, \cdots$ are negligibly small at  $eV_\text{AC}\ll\hbar\omega$.
A spike  appears at $2e V_\textrm{DC} / (\hbar \omega) = n / p$ when the $p$-th harmonics of the $n T_\textrm{J}$-periodic supercurrent exists in the DC limit of $V_\textrm{AC} = 0$, showing a signature of the $2n \pi$ fractional AC Josephson effect.
The Shapiro spikes can be well distinguished from the conventional case in which the main Shapiro spikes appear at $2e V_\textrm{DC} / (\hbar \omega) = 1/p$, since the spike positions $n/p$ (or the period-elongation factor $n$) depend on $V_\textrm{DC}$ in the $2 \pi n$ fractional AC Josephson effect. 

In this section, we study the Shapiro spikes of the $2n \pi$ fractional AC Josephson effect in the regime of the large AC voltage bias $eV_\textrm{AC} \sim \hbar \omega \lesssim 2e V_\textrm{DC}$. In this regime, the sub-spikes of $q = 2,3, \cdots$ become non-negligible so that the Shapiro spikes have a richer structure, revealing the dynamics of the fractional $2n \pi$ AC Josephson effects.
It is possible to experimentally explore this regime.
 
Below we derive the contribution of the Majorana zero modes to the Josephson current in the presence of $V_\textrm{AC}$ in Sec.~\ref{Subsec:Shapiro}.
The derivation is necessary in the regime of  the large AC voltage bias $eV_\textrm{AC} \sim \hbar \omega \lesssim 2e V_\textrm{DC}$, to find the positions of the sub-spikes of $q = 2,3, \cdots$;  the positions of the main Shapiro spikes are easily expected in the opposite regime of a lower AC voltage bias.
In Sec.~\ref{Subsec:SizeShapiro}, we compare the sub-spikes of the $2n\pi$ fractional AC Josephson effects with those of the conventional $2\pi$ supercurrent at large AC voltage bias. We also discuss the contribution of the midgap states to the Shapiro spikes.

\subsection{Shapiro spikes by Majorana zero modes}\label{Subsec:Shapiro}

We derive the contribution of the Majorana zero modes to the Josephson current when the AC voltage bias $V(t) = V_\textrm{DC} + V_\textrm{AC} \sin (\omega t)$ is applied across the junction.
%Here, we consider the Josephson effect under both DC and AC voltage bias with relatively large AC voltage $V_\text{AC}$ and find that the missing Shapiro steps can be identified with increased $V_\text{AC}$.
 The AC voltage results in the time dependence of the SC phase difference $\varphi_0$ as   $\varphi_0 (t) =\omega_\text{J}t + (\overline{\omega}/\omega)\sin\omega t + \phi_0$, where $\omega_\text{J}\equiv 2\pi/T_\text{J}$ and $\overline{\omega}\equiv2eV_\text{AC}/\hbar$. 
 
 We first consider the case of $n=2$ (the $2T_\text{J}$-periodic fractional Jospheson effect).
 In the limit of $V_\textrm{AC} =0$, the supercurrent $I_\text{MF}(t)$ contributed by the Majorana zero modes is derived in Eq.~\eqref{Eq:IMFn2-2}. Replacing the terms $2 \pi t / T_\textrm{J}$ in Eq.~\eqref{Eq:IMFn2-2} with the above expression of $\varphi_0(t)$, we obtain the supercurrent $I_\text{MF}(t)$ at nonzero $V_\textrm{AC}$,  
\begin{eqnarray*}
I_\text{MF}^{(n=2)}(t)=\frac{4eD_\delta}{\hbar}\sum_{p=1}^\infty pI_{2p}(\delta)
\left\{
J_0\left(\frac{p\overline{\omega}}{2\omega}\right)\sin\left(\frac{p\omega_\text{J}t}{2}+\phi_0\right) \right. \\
\left.
(-1)^s\sum_{q=1}^\infty\sum_{s=\pm1}J_q\left(\frac{p\overline{\omega}}{2\omega}\right)
\sin\left[\phi_0+\left(\frac{p\omega_\text{J}}{2}+sq\omega\right)t\right]
\right\},
\end{eqnarray*}
where $J_q$ is the Bessel function of the first kind. Only when the driving frequency of $\omega$ satisfies $\omega=\pm\frac{p\omega_\text{J}}{2q}$, the non-zero DC component arises with the intensities proportional to $pI_{2p}(\delta)J_q(p\overline{\omega}/(2\omega))$. 
In the regime of the weak AC voltage regime of  $eV_\text{AC}\ll \textrm{min} \{ 2e V_\textrm{DC}, \hbar\omega \}$, the results of the previous study~\cite{ChoiSim} are reproduced: The main Shaprio spikes appear at $\omega_\text{J}/\omega=\pm 2/p$ and the other spikes are negligible; the asymptotic form $J_q(x)\sim x^q/(2^q q!)$ at $x \lesssim 1$ was used.
%While with conventional $T_\text{J}$-periodic current, $I_\text{DC}$'s are found when $\omega_\text{J}/\omega=1$ under the regime $2eV_\text{AC}/(\hbar\omega)\ll1$, the $2T_\text{J}$-periodic current provides the non-vanishing $I_\text{DC}$ when $\omega_\text{J}/\omega=2$. Although the conventional case can provide $I_\text{DC}$ when $\omega_\text{J}/\omega=2$ with larger $2eV_\text{AC}/(\hbar\omega)\sim 1$, we will show that two cases can be distinguished by comparing the magnitude of $I_\text{DC}$ as a function of $V_\text{AC}$ in the subsection~\ref{Subsec:SizeShapiro}.
 
Similarly, we obtain the supercurrent $I_\text{MF}(t)$ at nonzero $V_\textrm{AC}$ in the case of $n=3$ (the $3T_\text{J}$-periodic fractional Jospheson effect),
\begin{eqnarray*}
\!I^{(n=3)}_\text{MF}(t)\!=\!\frac{4eD_{\frac{9\delta}{4}}}{\hbar}\sum_{p=1}^\infty pI_{2p}\!\left(\!\frac{9\delta}{4}\!\right)
\left\{\!
J_0\left(\!\frac{p\overline{\omega}}{3\omega}\!\right)\sin\left(\!\frac{p\omega_\text{J}t}{3}+\phi_0\!\right) \right. \\
\left.
(-1)^s\sum_{q=1}^\infty\sum_{s=\pm1}J_q\left(\frac{p\overline{\omega}}{3\omega}\right)
\sin\left[\phi_0+\left(\frac{p\omega_\text{J}}{3}+sq\omega\right)t\right]
\right\},
\end{eqnarray*}
%The dominating DC currents are found when $\omega_\text{J}/\omega = 3/p=3,1.5,1,\cdots$. The contrast to the conventional case is the missing $I_\text{DC}$ at $\omega_\text{J}/\omega = 2$. When the conventional case presents $I_\text{DC}$t at $\omega_\text{J}/\omega = 3$, the stronger $I_\text{DC}$'s show up at $\omega_\text{J}/\omega = 1, 2$ for $V_\text{DC}$. In addition, the missing component at $\omega_\text{J}/\omega = 2$ distinguishes the evolution of Majorana qubit by non-Abelian effect from single-electron tunneling which results in the missing component at $\omega_\text{J}/\omega = 1$. $I_\text{MF}(t)$ with $V_\text{DC}$ and $V_\text{AC}$ for $n=4$ is
and the case of $n=4$ (the $4T_\text{J}$-periodic fractional Jospheson effect),
\begin{eqnarray*}
I^{(n=4)}_\text{MF}(t)=\frac{2eD_\delta}{\hbar}\sum_{p=1 \atop p \text{ odd}}^\infty pI_{p}(\delta)
\left\{
J_0\left(\frac{p\overline{\omega}}{4\omega}\right)\sin\left(\frac{p\omega_\text{J}t}{4}+\phi_0\right) \right. \\
\left.
(-1)^s\sum_{q=1}^\infty\sum_{s=\pm1}J_q\left(\frac{p\overline{\omega}}{4\omega}\right)
\sin\left[\phi_0+\left(\frac{p\omega_\text{J}}{4}+sq\omega\right)t\right]
\right\}.
\end{eqnarray*}
The results of the previous study~\cite{ChoiSim} are also reproduced
in the regime of the weak AC voltage regime of  $eV_\text{AC}\ll \textrm{min} \{ 2e V_\textrm{DC}, \hbar\omega \}$.

%%%%%%%%%%%%%%%%%%%%%%%%%%%%%%%%%%%%% FIGURE.2 BEGIN %%%%%%%%%%%%%%%%%%%%%%%%%%%%%%%%%%%%%
\begin{figure*}[ht]
\centering
\includegraphics[width=1.8\columnwidth]{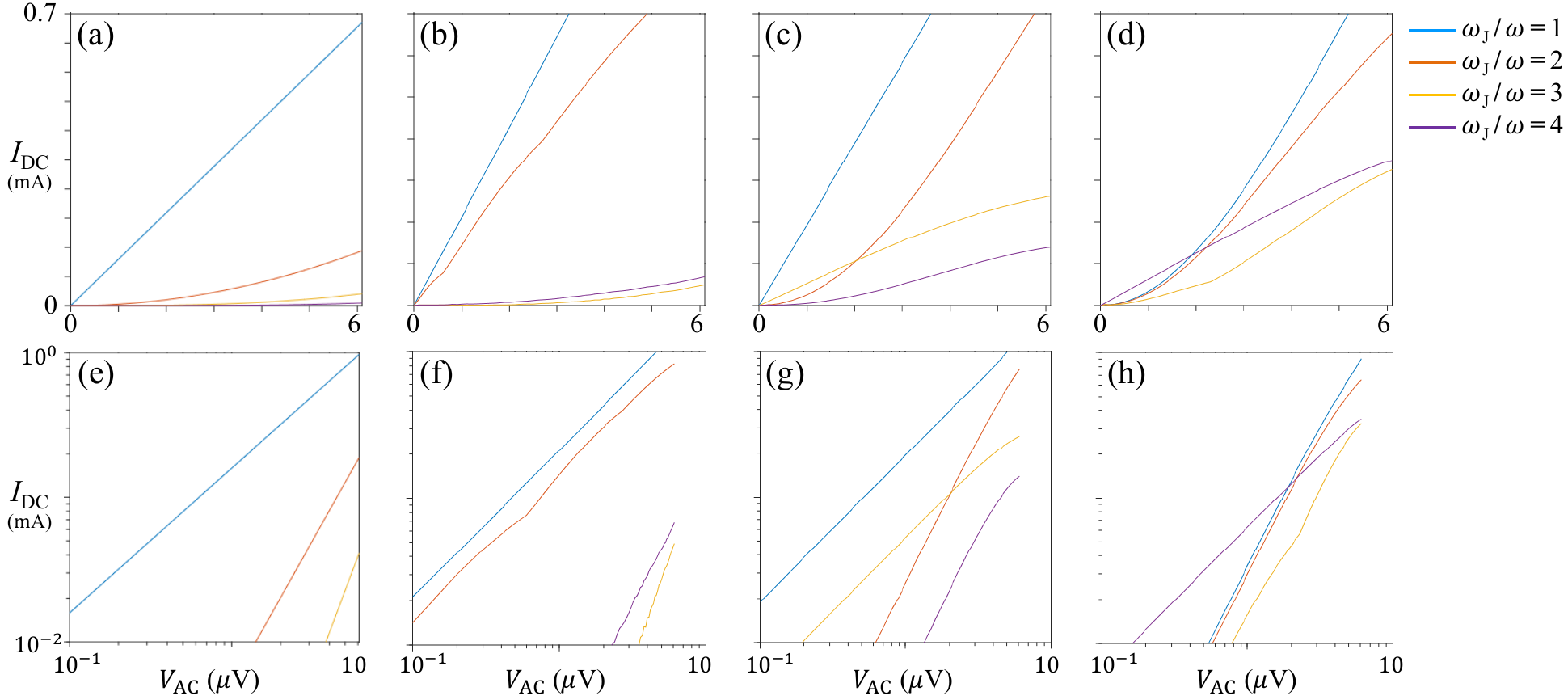}
\caption{Numerical calculations of the height $I_\textrm{DC}$ of the Shaprio spikes (generated by the Majorana zero modes) arising at $\omega_\text{J}/\omega=1,2,3,4$ when  the AC voltage bias $V(t) = V_\textrm{DC} + V_\textrm{AC} \sin (\omega t)$ is applied. 
The cases of  (a) the conventional $2\pi$ Josephson effect, (b) the $2 T_\textrm{J}$-periodic fractional AC Josephson effect,
(c) the $3 T_\textrm{J}$-periodic fractional effect, (d) the $4 T_\textrm{J}$-periodic fractional effect are shown.
The lower panels (e)-(h) show the log-log plot of the panels (a)-(d), respectively. 
At low $V_\textrm{AC}$ only the main spikes appear (the sub-spikes are negligilble), while the sub-spikes are also shown at high $V_\textrm{AC}$.
Among $\omega_\text{J}/\omega=1,2,3,4$, the positions of the main spikes are
$\omega_\text{J}/\omega = 1$ in the conventional case (a),
$\omega_\text{J}/\omega =1,2$ in the $2 T_\textrm{J}$-periodic fractional effect (b),
$\omega_\text{J}/\omega =1,3$ in the $3 T_\textrm{J}$-periodic fractional effect (c),
and
$\omega_\text{J}/\omega =4$ in the $3 T_\textrm{J}$-periodic fractional effect (d);
the other values of  $\omega_\text{J}/\omega$ are sub-spike positions.
The main spikes linearly increase with $V_\textrm{AC}$, as shown in the log-log plots, while the sub-spikes do not.
This property is useful for distinguishing between the main spikes and sub-spikes at large $V_\textrm{AC}$, hence,
useful for distinguishing between the $2n\pi$ fractional AC Josephson effect and the conventional $2 \pi$ Josephson effect.
% In conventional case, only $\log I_\text{DC}$ at $\omega_\text{J}/\omega=1$ has the slope $q=1$ as a function of $V_\text{AC}$ (e), while $2n\pi$ fractional AC Josephson effect has the slope of unity at higher $\omega_\text{J}/\omega$. In the $n=2$ case, $q=1$ occurs at $\omega_\text{J}/\omega=1$ and $\omega_\text{J}/\omega=2$ (f). In the $n=3$ case, $q=1$ occurs at $\omega_\text{J}/\omega=1$ and $\omega_\text{J}/\omega=3$ (g). In the $n=4$ case, $q=1$ occurs only at $\omega_\text{J}/\omega=4$ (h). 
The parameters suggested in Fig.~\ref{Fig:MFsN3} are used for the calculations.
} \label{Fig:eVac}
\end{figure*}
%%%%%%%%%%%%%%%%%%%%%%%%%%%%%%%%%%%%% FIGURE.2 END %%%%%%%%%%%%%%%%%%%%%%%%%%%%%%%%%%%%%

\subsection{Shapiro sub-spikes at large $V_\text{AC}$}\label{Subsec:SizeShapiro}

We compare the sub-spikes of the $2n\pi$ fractional AC Josephson effeects with those of the conventional $2\pi$ supercurrent at large AC voltage bias.  From the above expressions of  $I^{(n)}_\text{MF}(t)$,  we obtain the height of the Shapiro spikes (generated by the Majorana zero modes)  at $\omega_\text{J} / \omega=\pm nq / p $
in the $2n \pi$ fractional AC Josephson effect,
%On the other hand, the amplitudes of $I_\text{DC}$ at integer multiple of $\omega_\text{J}/\omega$ from $2n\pi$ fractional Josephson effect are
\begin{eqnarray}
I_\text{DC}\propto
\left\{
{\setlength\arraycolsep{1pt}
\begin{array}{l}
\sum\limits_{p=1}^{\infty}pI_{2p}\left(\frac{n^2\delta}{4}\right)J_{q}\left(\frac{p}{n}\frac{2eV_\text{AC}}{\hbar\omega}\right), \quad n=2,3, \\
\\
\sum\limits_{p=1\atop p\text{ odd}}^{\infty}pI_{p}\left(\delta\right)J_{q}\left(\frac{p}{n}\frac{2eV_\text{AC}}{\hbar\omega}\right), \quad\quad\,\, n=4.
\end{array}
}\right. \label{Eq:Idc}
\end{eqnarray}
The spike height corresponds to the size of Shapiro steps.~\cite{Tinkham}
%  Experimentally, our analysis on the amplitude of $I_\text{DC}$ can be done with the size of Shapiro steps.

At low $eV_\text{AC}\ll \textrm{min} \{ 2e V_\textrm{DC}, \hbar\omega \}$, the Shapiro spikes of the $2n \pi$ fractional AC Josephson effect appear at the positions $\omega_\text{J}/\omega$ different from those of the conventional $2\pi$ Josephson effect.
As mentioned above, the spikes of $q=1$ called the main spikes, as they appear at the low-$V_\textrm{AC}$ regime.
And, the other spikes of $q \ne 1$ is called sub-spikes as they are negligible at the regime.
The main spikes appear at $\omega_\textrm{J} / \omega = \pm n / p$ with $p = 1,2,3 \cdots$ for $n=2,3$ and $p = 1,3,5, \cdots$ for $n=4$ in the $2n \pi$ fractional Josephson effect. We notice that the main spikes appear not at every integer value of $\omega_\textrm{J} / \omega$; namely, there are missing Shapiro spikes as in the known $4 \pi$ fractional Josephson effect.~\cite{Wiedenmann,Furdyna,Molenkamp,Laroche} By contrast, the main spikes appear only at $\omega_\textrm{J} / \omega = \pm 1$ in the conventional $2 \pi$ Josephson effect. Therefore, the $2n \pi$ fractional AC Josephson effect can be well distinguished from the conventional $2 \pi$ Josephson effect by observing Shapiro spikes as pointed out in Ref.~\cite{ChoiSim}.

We point out that the dependence of the missing components on $V_\text{DC}$ cannot be reproduced by other accidental artefacts~\cite{Egger,DasSarma,Mora}; the known  $4 \pi$ fractional Josephson effect~\cite{Wiedenmann,Furdyna,Molenkamp,Laroche} can be reproduced by such artefacts.
It is because the $2n \pi$ fractional AC Josephson effect originates from the non-Abelian state evolution.
The non-Abelian effect of the Majorana zero modes is well distinguished from the artefacts.

%$I_\text{DC}$ emerge when $\omega_\text{J}/\omega = 4/p=4,4/3,4/5,\cdots$, resulting in more missing $I_\text{DC}$ at $\omega_\text{J}/\omega = 1,2,3$. Similarly to $n=3$ case, the missing components distinguish the evolution of Majorana qubit by non-Abelian effect from single-electron tunneling. Moreover, we can compare the missing components for $n=3$ and $n=4$ cases by tuning $V_\text{DC}$ with an identical sample, and

%In conventional $2\pi$ Josephson effects, $I_\text{DC}$ arises with integer multiple of $\omega_\text{J}/\omega=q$ whose amplitude is proportional to $J_q(2eV_\text{AC}/(\hbar\omega))$~\cite{Tinkham} [see Fig.~\ref{Fig:eVac}(a) and (e)]. 
  
%As shown in the subsection~\ref{Subsec:Shapiro}, we can distinguish, by the missing $I_\text{DC}$, the $2n\pi$ fractional AC Josephson effect from the conventional AC Josephson effects, while the previous subsection assumes $2eV_\text{AC}/(\hbar\omega)\ll1$ and neglects $q\ge2$ cases which can fill up the missing $I_\text{DC}$ in accidental combinations with $p$-th high harmonics; $I_\text{DC}$ arises when $\omega_\text{J}/\omega=\pm nq/p$ for $2n\pi$ fractional Josephson effect.

On the other hand, as $V_\textrm{AC}$ increases, sub-spikes (including the missing Shapiro spikes) become pronounced.
The sub-spikes are still distinguishable from the main spikes even at high $eV_\text{AC}\sim\hbar\omega$. 
%We note that missing $I_\text{DC}$ are exhibited with $q=1$ (subsection~\ref{Subsec:Shapiro}), and the missing $I_\text{DC}$ can be filled up in combination with $q\ge2$ and $p$-th high harmonics. Hence, if we can recognize $q$ from the $I_\text{DC}$ arising at an integer of $nq/p$, we can discern whether it is the missing $I_\text{DC}$. In order that, let us analyze $\log I_\text{DC}$ as a function of $\log V_\text{AC}$. 
From Eq.~\eqref{Eq:Idc}, we find   
\begin{equation}
\log I_\text{DC} \propto \text{const.} + q\log V_\text{AC},
\end{equation}
using the asymptotic formula of $J_q(x)\sim x^q/(2^q q!)$  at $x \lesssim 1$.
%working with 3 percent of maximum deviation from $J_q(x)$ within $0\le x\le1/2$,
This shows that the height $I_\textrm{DC}$ of the main spikes, which have $q=1$, linearly increases with $V_\textrm{AC}$.
The value of $q$ can be obtained from the log-log plot of the height versus $V_\textrm{AC}$ (see Fig.~\ref{Fig:eVac}).
This property is useful for distinguishing between  the main spikes and sub-spikes at large $V_\textrm{AC}$ (hence,
useful for distinguishing between the $2n\pi$ fractional AC Josephson effect and the conventional $2 \pi$ Josephson effect)
in the experimental situation where the low-$V_\text{AC}$ regime is not achievable.

The analytic expression of the spike height in Eq.~\eqref{Eq:Idc} is in good agreement with
the numerical result obtained by solving the Hamiltonain $H$ in Eq.~\eqref{Eq:Heff}.
The numerical result is shown in  Fig.~\ref{Fig:eVac}(b)-(d) for $n=2,3,4$ cases. 

%In the regime $2eV_\text{AC}/(\hbar \omega)\ll1$, the most dominating $I_\text{DC}$ appear as analysed in the previous subsection~\ref{Subsec:Shapiro}, while $I_\text{DC}$ at $\omega_\text{J}/\omega=1,2,3,4$ begin to appear and fill up the missing $I_\text{DC}$ as increasing $V_\text{AC}$.

% In this subsection, we analyze the amplitudes of $I_\text{DC}$ arising at integer multiple of $\omega_\text{J}/\omega$  as a function of $V_\text{AC}$, and the contrast of $2n\pi$ fractional Josephson effect to the known effects becomes more evident. We show that the parasitic artefacts filling up the missing $I_\text{DC}$ can be distinguished by inspecting the dependence on $V_\text{AC}$. We compare the analysis to the numerical calculation of $I_\text{DC}$ which is  evaluated from the time evolution of Majorana qubit with brute force.

%We present the numerical calculation of $\log I_\text{DC}$ as a function of $\log V_\text{AC}$ in Fig.~\ref{Fig:eVac}(e)-(h). All missing $I_\text{DC}$ occur with $q\ge2$. Even when the missing $I_\text{DC}$ under high $V_\text{AC}$ can be stronger than $I_\text{DC}$ persisting at $2eV_\text{AC}\ll \hbar\omega$, we can discern the missing $I_\text{DC}$.

We discuss the contribution of the midgap states to the height of the Shapiro spikes.
Since the midgap states result in the $2 \pi$ periodic supercurrent, they do not modify the Shapiro spikes at
$\omega_\text{J} / \omega > 1$ in the low-$V_\textrm{AC}$ regime; the midgap states affect the Shapiro spikes only at $\omega_\text{J} / \omega = 1/p$ with $p = 1,2, \cdots$.
Similarly, the midgap states do not modify the main spikes except the trivial main spike at $\omega_\text{J} / \omega = 1$ in the high-$V_\textrm{AC}$ regime.
Therefore, the characteristic features of the $2n\pi$ fractional AC Josephson effect are maintained even in the presence of the midgap states.

%$I_\text{DC}$ of $I_\text{MF}(t)$ and find that the missing component of $n=3$ and $n=4$ cases are not affected from the midgap states. Note that midgap states result in the $2\pi$-periodic Josephson current since they are energetically well-separated from the Majorana fermions which exhibit energy level crossing and non-Abelian effect. Accordingly, $I_\text{MF}(t)$ under $V_\text{DC}$ consists of complex combinations of frequency components of $p\omega_\text{J}$ with higher harmonics, and $I_\text{DC}$ of $I_\text{MF}(t)$ under both $V_\text{DC}$ and $V_\text{AC}$ arise at $\omega_\text{J}/\omega=1/p$ in the regime $\overline{\omega}/\omega\ll1$, where $p=1,2,3,\cdots$. Hence, $I_\text{DC}$ from midgap state does not fill up all the missing $I_\text{DC}$ of $\omega_\text{J}/\omega=2$ and $\omega_\text{J}/\omega=2,3$ of the $6\pi$ and $8\pi$ fractional AC Josephson effect, respectively. Even in exotic $2\pi$-periodic Josephson junctions showing non-sinusoidal current phase relation, it does not exhibit $I_\text{DC}$ at $\omega_\text{J}/\omega\ge 2$ with $2eV_\text{AC}/(\hbar \omega)\ll 1$. Even though $I_\text{DC}$ of $I_\text{MF}(t)$ in those systems can emerge at integer multiples of $\omega_\text{J}/\omega=1/p$ with $2eV_\text{AC}/(\hbar\omega)\sim 1$ so that $\omega_\text{J}/\omega\ge 2$, the missing $I_\text{DC}$ do not exist.

\section{Josephson junction with $N \ne 3$ or with Abrikosov vortices}\label{Sec:MultipleMFs}

In Ref.~\cite{ChoiSim}, the $2n \pi$ fractional AC Josephson effect was predicted, focusing on the case that $N=3$ magnetic flux quanta is applied to the junction and that there is no Abrikosov vortex inside the SC regions. In this case, the fusion and braiding of four Majorana zero modes can happen as discussed in Sec.~\ref{Subsec:FJ}. 
Below, we study the evolution of Majorana zero modes as a function of the superconducting phase difference across the junction in the case of $N=1,2,4$ (in the absence of Abrikosov vortices) in Sec.~\ref{Subsec:withoutVortices}, generalizing the case of $N=3$. We find an even-odd effect of $N$. For odd $N$, the fusion of Majorana zero modes happens, resulting in nonzero Josephson current $I_\text{MF}$. For even $N$, no fusion appears and the junction does not exhibit any Josephson current as in the case of the conventional Josephson junction with an integer number of magnetic flux quanta (i.e., at the nodes of the conventional Fraunhofer pattern~\cite{Tinkham}). In the case of $N=4$, braiding of four Majorana zero modes can happen, although it is not accompanied by fusions.  

In Sec.~\ref{Subsec:withVortices}, we also study the effects of the Abirokosov vortices on the $2 n \pi$ fractional AC Josephson effect in the case of $N=2,3$. We find that a modified $2n \pi$ fractional Josephson effect occurs with any number of Abrikosov vortices, if a suitable $N$ is chosen.

%%%%%%%%%%%%%%%%%%%%%%%%%%%%%%%%%%%%%% FIGURE BEGIN %%%%%%%%%%%%%%%%%%%%%%%%%%%%%%%%%%%%%
%\begin{figure*}[ht]
%%\centering
%%\includegraphics[width=0.85\linewidth]{prbMFsN1N2.pdf}
%\includegraphics[width=1.56\columnwidth]{prbMFsN1N2.pdf}
%\caption{(Color online) Midgap state spectrum are depicted as a function of $\varphi_0$ under $N=1$ and $N=2$ in (a) and (c), respectively. Note that the complex fermion state with quasiparicle $|1\rangle$ has the positive excitation spectrum.  The spatial probability density of Majorana fermions are drawn similarly as in Fig.~\ref{Fig:MFsN3} with the same parameters under $N=1$ and $N=2$ in (b) and (d), respectively.
%} \label{Fig:MFsN1N2}
%\end{figure*}
%%%%%%%%%%%%%%%%%%%%%%%%%%%%%%%%%%%%%% FIGURE END %%%%%%%%%%%%%%%%%%%%%%%%%%%%%%%%%%%%% 
%%%%%%%%%%%%%%%%%%%%%%%%%%%%%%%%%%%%% FIGURE BEGIN %%%%%%%%%%%%%%%%%%%%%%%%%%%%%%%%%%%%%
\onecolumngrid
\begin{center}
\begin{figure}[t]
%\centering
%\includegraphics[width=0.85\linewidth]{prbMFsN1N2.pdf}
\includegraphics[width=0.9\columnwidth]{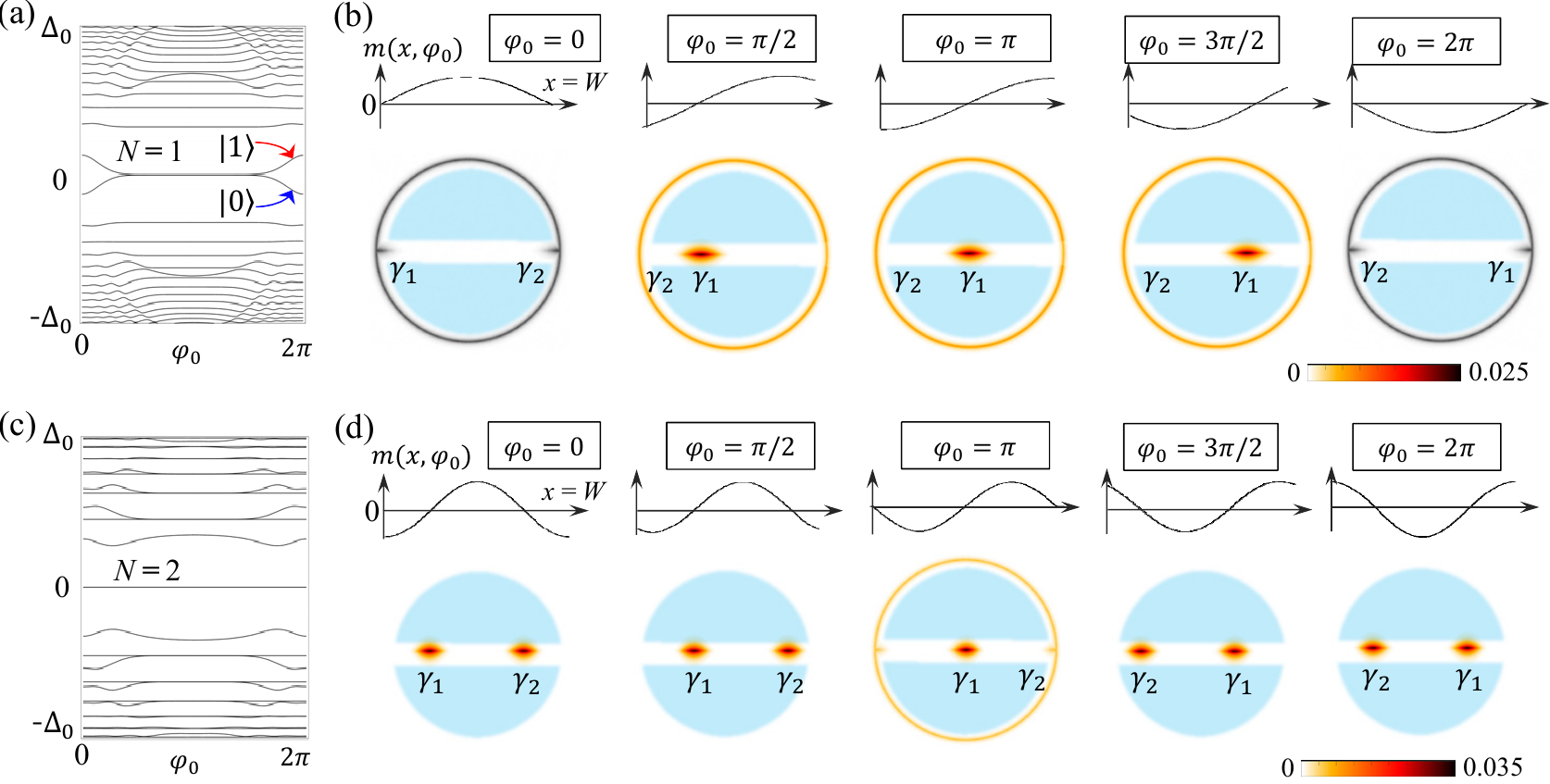}
\caption{ (a,c) Numerical results of single-particle energy levels of the setup in the cases of (a) $N=1$ and (c) $N=2$, as a function of $\varphi_0$. In the case of $N=1$, fusion between two Majorana zero modes happens in a range of $\varphi_0$, resulting in nonzero energies as marked by arrows. 
(b,d) The spatial probability density of Majorana fermions is drawn at various $\varphi_0$ as in Fig.~\ref{Fig:MFsN3}.
The results of this figure are obtained with the parameters used in Fig.~\ref{Fig:MFsN3}, excepts $N=1$ in (a,b) and $N=2$ in (c,d).}
 \label{Fig:MFsN1N2}
\end{figure}
\end{center}
\twocolumngrid
%%%%%%%%%%%%%%%%%%%%%%%%%%%%%%%%%%%%% FIGURE END %%%%%%%%%%%%%%%%%%%%%%%%%%%%%%%%%%%%%% 

\subsection{Josephson junction in the case of $N=1,2,4$}\label{Subsec:withoutVortices}

We first consider the case of $N=1$ and $N=2$ in the absence of Abrikosov vortices.
The emergence of Majorana fermions is found in Fig.~\ref{Fig:MFsN1N2}, by numerically diagonalizing $H$ in Eq.~\eqref{Eq:Heff}.
At $\varphi_0 = \pi$, there appear one localized Majorana zero mode at the center of the junction and an extended chiral Majorana zero mode propagating along the arcs in both the cases of $N=1$ and $N=2$. 
The presence of the extended chiral Majorana zero mode satisfies the quantization rule in Eq.~\eqref{Eq:QuantRule}.

The evolution of the two Majorana fermions, as a function of the superconducting phase difference $\varphi_0$ across the junction, depends on whether $N$ is 1 or 2.
As $\varphi_0$ increases, the localized Majorana zero mode moves towards the right end of the junction.
In the $N=1$ case, there are still one localized Majorana zero modes inside the junction and the extended chiral Majorana zero mode along the arcs at $\varphi_0 = 3 \pi/2$.
By contrast, in the $N=2$ case, there are two localized Majorana zero modes inside the junction at $\varphi_0 = 3 \pi/2$;
the extended chiral Majorana zero mode at $\varphi=\pi$ now moves fully inside the junction; there is no extended chiral Majorana zero mode any more at $\varphi_0 = 3 \pi/2$ in this case of $N=2$.
The difference between $N=1$ and $N=2$ can be understood from the fact that the separation $W/N$ between two adjacent localized Majorana fermions is shorter for $N=2$ as in Eq.~\eqref{Eq:positionMF}; namely, the junction can host two localized Majorana zero modes at certain $\varphi_0$ for $N=2$, while it can host at most one localized Majorana zero modes for $N=1$.

In the $N=1$ case, the localized zero mode approaches sufficiently close to the junction end at larger $\varphi_0$,
and there occurs fusion between the localized Majorana zero mode and the extended chiral Majorana zero mode [see the case of $\varphi_0 = 2\pi$ in Fig.~\ref{Fig:MFsN1N2}(b)]. 
The fusion (or the collison between the localized zero mode in the junction and the extended zero mode along the arcs) must occur in a certain range of $\varphi_0$, since the localized Majorana zero mode moves by the distance that equals the junction width $W$ while $\varphi_0$ changes by $2\pi$.
The fusion results in non-zero single-particle energies as shown in Fig.~\ref{Fig:MFsN1N2}(a),
and generates nonzero Josephson current, contrary to the conventional Josephson junction at $N=1$ magnetic flux quantum. 
As $\varphi_0$ increases further, the two Majorana fermions split into one localized Majorana zero mode inside the junction and the other extended chiral Majorana zero mode along the arc [see the case of $\varphi_0 = \pi/2$ in Fig.~\ref{Fig:MFsN1N2}(b)].

On the other hand, the $N=2$ case  does not exhibit  any fusion, since the two Majorana zero modes are spatially well separated at any $\varphi_0$ [see  Fig.~\ref{Fig:MFsN1N2}(d)].
%; while $\varphi_0$ changes by $2 \pi$, a localized Majorana zero mode moves by the distance $W/2$ that is half of the junction width $W$.
As a result, the junction has the zero-energy single-particle state at any $\varphi_0$  [see  Fig.~\ref{Fig:MFsN1N2}(c)],
and no Josephson current, similarly to the conventional Josephson junction at $N=2$ magnetic flux quanta.
Note that our numerical computation of the Josephson current (including the midgap states) confirms this.

%  When $N=1$ and $\varphi_0=0$, no Majorana fermions appear along the junction and arcs. As $\varphi_0$ increases, a localized Majorana fermions appears in the junction while leaving a chiral Majorana fermion surrounding arcs, and the Majorana fermions disappear and fuse into a complex fermion state [see Fig.~\ref{Fig:MFsN1N2}(b)]. Under $N=1$, the total number of Majorana fermion changes by the fusion of two Majorana fermions since a localized Majorana fermion emerges in the junction; this is in accordance with Eq.~\eqref{Eq:QuantRule}. On the other hand, When $N=2$ and $\varphi_0=0$, two localized Majorana fermions appear in the junction. As $\varphi_0$ increases, one of the Majorana fermions in the junction becomes the chiral Majorana fermion along the arcs. Under $N=2$, the total number of Majorana fermions remains the same and no fusion happens.  In total, $I_\text{MF}$ and the fusion of Majorana fermions occur when $N=1$ contrary to conventional Josephson junctions in which the Josephson current vanishes with the integer number of $N$. When $N=2$, none of $I_\text{MF}$ and fusion arise. In addition, numerical computations of the Josephson current including midgap states yield the same result.

%%%%%%%%%%%%%%%%%%%%%%%%%%%%%%%%%%%%%% FIGURE BEGIN %%%%%%%%%%%%%%%%%%%%%%%%%%%%%%%%%%%%%
\begin{figure*}[!t]
\centering
\includegraphics[width=1.8\columnwidth]{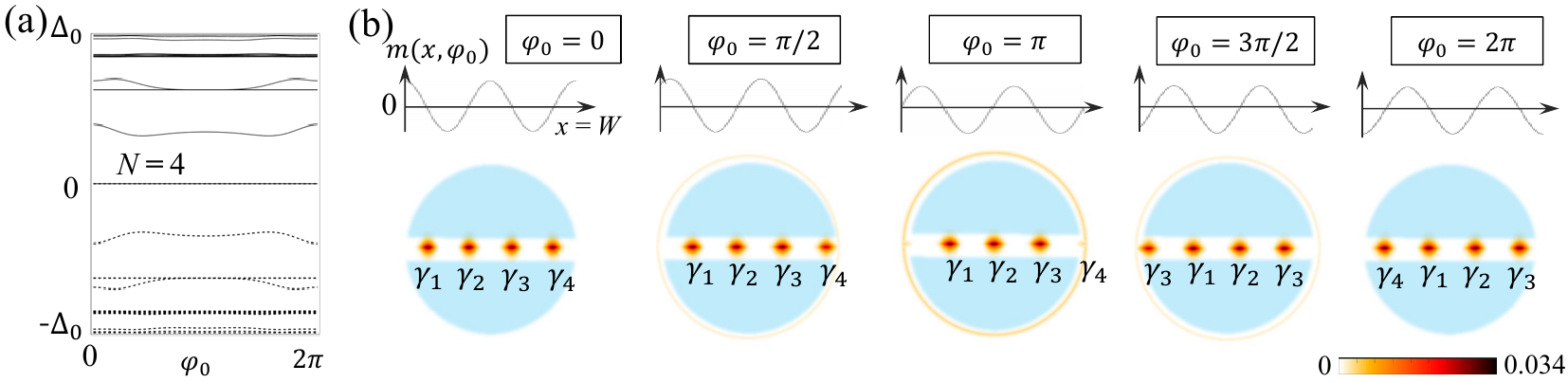}
\caption{ (a) Numerical results of single-particle energy levels of the setup in the $N=4$ case, as a function of $\varphi_0$. In this case, the junction hosts four Majorana zero modes and does not show fusion between the zero modes at any $\varphi_0$. As a result, there are always two-fold degenerate zero-energy states at any $\varphi_0$.
(b) The spatial probability density of Majorana fermions is drawn at various $\varphi_0$ as in Fig.~\ref{Fig:MFsN3}.
The results of this figure are obtained with the parameters used in Fig.~\ref{Fig:MFsN3}.
} \label{Fig:MFsN4}
\end{figure*}
%%%%%%%%%%%%%%%%%%%%%%%%%%%%%%%%%%%%%% FIGURE END %%%%%%%%%%%%%%%%%%%%%%%%%%%%%%%%%%%%% 
%%%%%%%%%%%%%%%%%%%%%%%%%%%%%%%%%%%%% FIGURE BEGIN %%%%%%%%%%%%%%%%%%%%%%%%%%%%%%%%%%%%%
%\onecolumngrid
%\begin{center}
%\begin{figure}[pt]
%\centering
%\includegraphics[width=0.8\columnwidth]{Fig8MFsN4.pdf}
%\caption{(Color online)  (a) Numerical results of single-particle energy levels of the setup in the $N=4$ case, as a function of $\varphi_0$. In this case, the junction hosts four Majorana zero modes and does not show fusion between the zero modes at any $\varphi_0$. As a result, there are always two-fold degenerate zero-energy states at any $\varphi_0$. (b) The spatial probability density of Majorana fermions is drawn at various $\varphi_0$ as in Fig.~\ref{Fig:MFsN3}. The results of this figure are obtained with the parameters used in Fig.~\ref{Fig:MFsN3}.
%} \label{Fig:MFsN4}
%\end{figure}
%\end{center}
%\twocolumngrid
%%%%%%%%%%%%%%%%%%%%%%%%%%%%%%%%%%%%% FIGURE END %%%%%%%%%%%%%%%%%%%%%%%%%%%%%%%%%%%%% 

The features of the $N=1$ and $N=2$ cases are generalized to general $N$, showing an even-odd behavior:
For odd $N$, fusion between a localized Majorana zero mode inside the junction and the extended chiral Majorana zero mode along the arcs occurs in a range of $\varphi_0$, while for even $N$ no fusion occurs at any $\varphi_0$.
This can be understood as follows.
According to Eq.~\eqref{Eq:positionMF}, the number of localized Majorana zero modes inside the junction for a given $\varphi_0$ and a given (even or odd) $N$ equals the number of the integers $k$ satisfying
\begin{equation}
\frac{-N+1}{2}-\frac{\varphi_0}{2\pi} < k < \frac{N+1}{2}-\frac{\varphi_0}{2\pi}. \label{Eq:NumMFs}
\end{equation}
For odd $N$ and at $\varphi_0 = \pi$, there are $N$ localized Majorana zero modes inside the junction and one extended chiral Majorana zero mode along the arcs. The total number of Majorana zero modes is $N+1$ which is even.
For odd $N$ and at $\varphi_0 = 2 \pi$ (or 0), there are $N-1$ localized Majorana zero modes inside the junction, while there is no extended chiral Majorana zero mode along the arcs. The total number $N-1$ of Majorana zero modes at $\varphi_0 = 2 \pi$ (or 0) is even, and smaller than
the number $N+1$ of Majorana zero modes at $\varphi_0 = \pi$ by two. The difference of the total number is due to the fusion.
The fusion results in the change of the energy of a complex fermion formed by the two fused Majorana fermions as a function of $\varphi_0$,  hence, generates nonzero Josephson current.

On the other hand, for even $N$ and at $\varphi_0 = \pi$, there are $N-1$ localized Majorana zero modes inside the junction and one extended chiral Majorana zero mode along the arcs. The total number of Majorana zero modes is $N$ which is even.
For even $N$ and at $\varphi_0 = 2 \pi$, there are $N$ localized Majorana zero modes inside the junction, while there is no extended chiral Majorana zero mode along the arcs. The total number of Majorana zero modes is always $N$ (and there is no fusion and no Josephson current) at any $\varphi_0$ for even $N$.

%From this condition, it can be shown that increasing $\varphi_0$ from 0 to $\pi$ causes the number of localized Majorana fermions to  increases (decreases) for the odd (even) number of $N$. We note that the chiral Majorana fermions and odd number of localized Majorana fermions and appears in the junction when $\varphi_0=\pi$ as $m(x,\varphi_0=\pi)=-m(W-x,\varphi_0=\pi)$, while the chiral Majorana fermion disappears when $\varphi_0=0$ as $m(x,\varphi_0=0)=m(W-x,\varphi_0=0)$. Considering the change of the number of Majorana fermions in the junction and surrounding the arcs, we see that the total number of Majorana fermions changes in pair only with the odd number of $N$. Consequently, the Josephson effect with $I_\text{MF}$ occurs only with the odd number of $N$ without the Abrikosov vortices. However, we point out that no non-Abelian effects in $I_\text{MF}$ can be exhibited with two Majorana fermions when $N=1,2$ due to the fermion parity conservation.

We point out that there is no non-Abelian evolution of the state of the junction in the cases of $N=1,2$, since
there are at most two Majorana zero modes in the system.
For $N \ge 3$, non-Abelian effects can happen. The case of $N=3$ was studied in Ref.~\cite{ChoiSim}.
We below show that the junction with $N=4$ exhibits non-Abelian state evolution although there occurs no fusion between Majorana zero modes. In the $N=4$ case,  we plot the numerical results of the single-particle energy spectrum of the junction and the probability density of Majorana zero modes in Fig.~\ref{Fig:MFsN4}. The junction has four Majorana zero modes at any $\varphi_0$.
We find that as $\varphi_0$ changes by $2 \pi$, the four Majorana zero modes are braided, following $\gamma_{i=1,2,3}\mapsto\gamma_{i+1}$ and $\gamma_4\mapsto-\gamma_1$  (here we use the notation for the four Majorana fermions used in Sec.~\ref{Subsec:FJ}).
This braiding operation is identical to the case of $N=3$ [see Eq.~\eqref{Eq:braiding}], but there is the crucial difference from $N=3$ that the braiding is not accompanied by any fusion between the zero modes in this case of $N=4$.
The braiding results in the state rotation in the ground-state manifold,
\begin{equation}
 |0_{41} 0_{32} \rangle_0 \mapsto \frac{1}{\sqrt{2}} (  |0_{41} 0_{32} \rangle_0 +  |1_{41} 1_{32} \rangle_0 ),  \label{Eq:evolutionN4}
 \end{equation}
 when the four Majorana zero modes form an initial state $|0_{41} 0_{32} \rangle_0$.
 No dynamical phase is attached, since the four Majorana fermions remain as zero modes at any $\varphi_0$. The state evolution is different from the $N=3$ case shown in Eq.~\eqref{Eq:Ue}.

%%%%%%%%%%%%%%%%%%%%%%%%%%%%%%%%%%%%%%% FIGURE BEGIN %%%%%%%%%%%%%%%%%%%%%%%%%%%%%%%%%%%%%
\begin{figure*}[pt]
\centering
\includegraphics[width=0.73\linewidth]{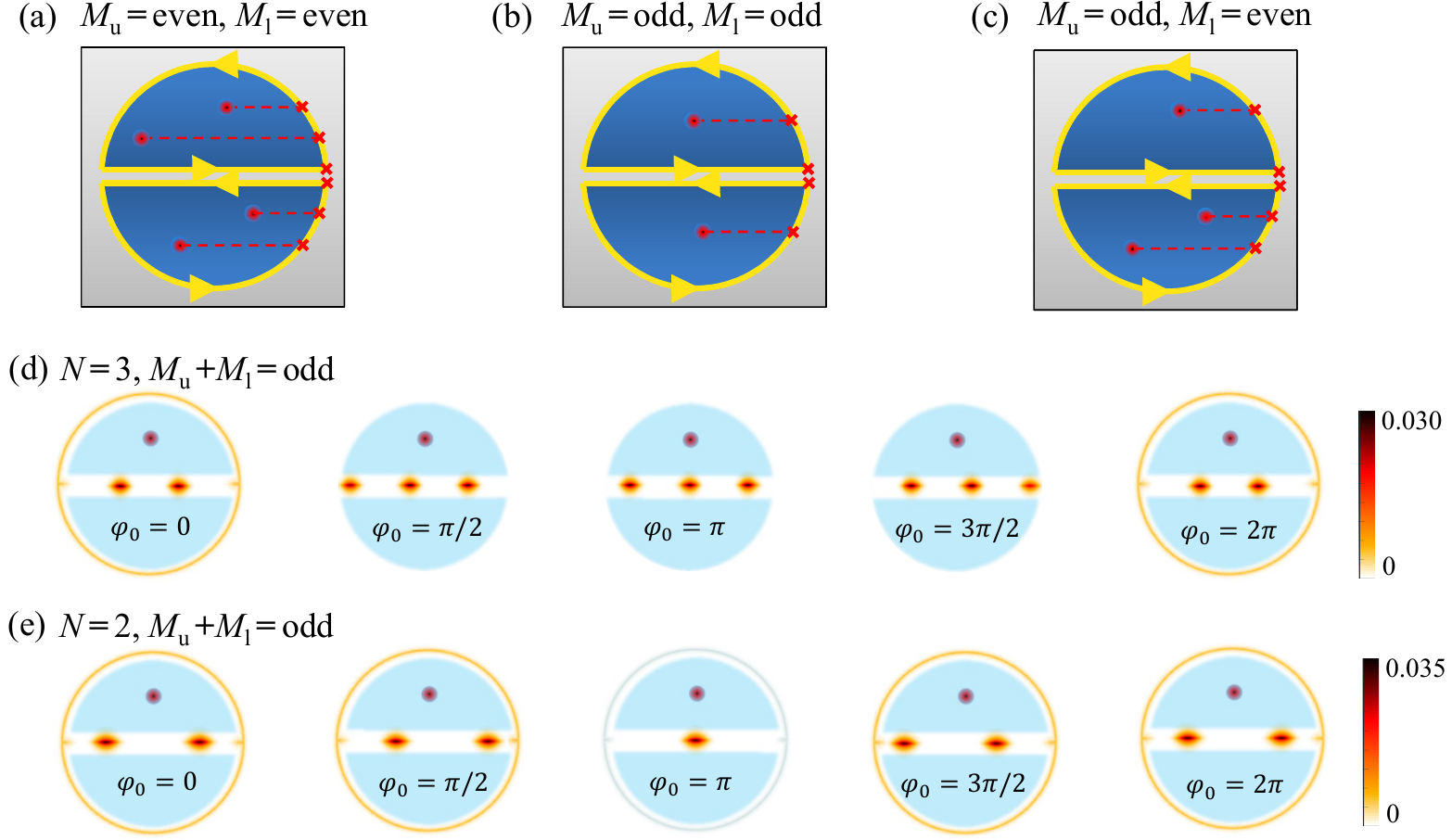}
\caption{(a-c) Different combinations of the parities of the number $M_\text{u}$ and $M_\text{l}$ of Abrikosov vortices in the SC regions. The red dots represent the Abrikosov vortices, and the red dashed lines show their branch cuts. The red crosses refer to the positions where the wave function of the chiral mode $\eta_{a=\textrm{u,l}}$ changes its sign. The combination of  even $M_\text{u}$ and odd $M_\text{u}$, which is not shown here, generates the same effect with the combination in (c). (d,c) The spatial probability density of Majorana fermions are drawn at $\varphi_0=0,\pi/2,\pi,3\pi/2,2\pi$ (from the left panels)
for (d) $N=3$ and odd $M_\textrm{u} + M_\textrm{l}$
and (e) $N=2$ and odd $M_\textrm{u} + M_\textrm{l}$ as in Fig.~\ref{Fig:MFsN1N2} with the same parameters. In the case of $N=3$, fusion of Majorana zero modes happens as $\varphi_0$ changes, while there is no fusion in the $N=2$ case.
} \label{Fig:MFsVortices}
\end{figure*}
%%%%%%%%%%%%%%%%%%%%%%%%%%%%%%%%%%%%%% FIGURE END %%%%%%%%%%%%%%%%%%%%%%%%%%%%%%%%%%%%%  
%%%%%%%%%%%%%%%%%%%%%%%%%%%%%%%%%%%%%% FIGURE BEGIN %%%%%%%%%%%%%%%%%%%%%%%%%%%%%%%%%%%%%
%\onecolumngrid
%\begin{center}
%\begin{figure}[ht]
%\centering
%\includegraphics[width=0.82\linewidth]{Fig9MFsVortices.pdf}
%\caption{(a-c) Different combinations of the parities of the number $M_\text{u}$ and $M_\text{l}$ of Abrikosov vortices in the SC regions. The red dots represent the Abrikosov vortices, and the red dashed lines show their branch cuts. The red crosses refer to the positions where the wave function of the chiral mode $\eta_{a=\textrm{u,l}}$ changes its sign. The combination of  even $M_\text{u}$ and odd $M_\text{u}$, which is not shown here, generates the same effect with the combination in (c). (d,c) The spatial probability density of Majorana fermions are drawn at $\varphi_0=0,\pi/2,\pi,3\pi/2,2\pi$ (from the left panels) for (d) $N=3$ and odd $M_\textrm{u} + M_\textrm{l}$ and (e) $N=2$ and odd $M_\textrm{u} + M_\textrm{l}$ as in Fig.~\ref{Fig:MFsN1N2} with the same parameters. In the case of $N=3$, fusion of Majorana zero modes happens as $\varphi_0$ changes, while there is no fusion in the $N=2$ case.
%} \label{Fig:MFsVortices}
%\end{figure}
%\end{center}
%\twocolumngrid
%%%%%%%%%%%%%%%%%%%%%%%%%%%%%%%%%%%%% FIGURE END %%%%%%%%%%%%%%%%%%%%%%%%%%%%%%%%%%%%%    

The braiding of the four Majorana fermions in the $N=4$ case cannot be detected by the Josephson current, differently from the $N=3$ case, since no Josephson current arises. We sketch a scheme for detecting the braiding. The scheme is composed of three steps.
As the first step, an initial state $|0_{41} 0_{32} \rangle_0$ is prepared, by adiabatically turning the magnetic field on from $N=0$ to $N=4$. In this process, the four Majorana zero modes $\gamma_{i=1,2,3,4}$ are nucleated. 
In the second step, the Majorana zero modes are braided by applying the DC voltage bias $V_\textrm{DC}$ during the time period of $T_\text{J}=h/(2eV_\text{DC})$ for which $\varphi_0$ changes by $2 \pi$.
The braiding results in the state rotation in Eq.~\eqref{Eq:evolutionN4}.
 In the last step, the magnetic field is adiabatically decreased from $N=4$ to $N=2$. Then the separation between the localized Majorana zero modes inside the junction increases from $W/4$ to $W/2$. As a result, two of the four Majorana fermions moves to the arcs of the junction and fuse. The fusion can genereate additional contribution to the Josephson current across the junction (in comparison with the case where the DC voltage bias is not applied) as the magnetic field decreases to $N=2$.
The additional contribution might be an evidence of the non-Abelian state evolution,
 and be detected by comparing the magnetic-field dependence of the Josephson current between the two cases with and without the application of the DC voltage bias.

\subsection{Effects of the Abrikosov vortices on the $2n \pi$ fractional AC Josephson effect}\label{Subsec:withVortices}

We consider the effect of Abrikosov vortices on the emergence of Majorana zero modes in the junction and on the $2n\pi$ fractional Josephson effect. The effect differs among the three cases of (i) even $M_\textrm{u}$ and even $M_\textrm{l}$,
(ii) odd $M_\textrm{u}$ and odd $M_\textrm{l}$, and (iii)  odd $M_\textrm{u}$ and even $M_\textrm{l}$ (equivalently,
even $M_\textrm{u}$ and odd $M_\textrm{l}$) [see Fig.~\ref{Fig:MFsVortices}(a)-(c)], where $M_\textrm{u}$ and $M_\textrm{l}$ are the number of Abrikosov vortices in the upper and lower SC regions, respectively.
We find that the $2n\pi$ fractional AC Josephson effect occurs with proper $N$ in the presence of the Abrikosov vortices;   the three cases (i)-(iii) have the different value of the proper $N$.

First, we consider the $N=3$ case and discuss how the Majorana zero modes in Abrikosov vortices affect the emergence of Majorana zero modes in the junction and arcs. When $M_\text{u}+M_\text{l}$ is even [see Fig.~\ref{Fig:MFsVortices}(a)-(b)], the emergence of the Majorana zero modes and their fusion are identical to the case of  $N=3$ and $M_\text{u}=M_\text{l}=0$ in the absence of the Abrikosov vortices. This can be understood from the quantization rule in Eq.~\eqref{Eq:QuantRule}. 

However, when $M_\text{u}+M_\text{l}$ is odd [see Fig.~\ref{Fig:MFsVortices}(c)], the total number of the Majorana zero modes (the localized zero modes inside the junction and the extended zero mode along the arcs) changes to be odd [see Fig.~\ref{Fig:MFsVortices}(d)]. When there are three localized Majorana zero modes inside the junction (for example at $\varphi_0 = \pi$), no extended zero mode appears along the arcs, contrary to the $N=3$ case in the absence of the Abrikosov vortices. When there are two localized Majorana zero modes inside the junction (at $\varphi_0 = 2 \pi$), there appears the extended chiral Majorana zero mode along the arcs.
The absence and appearance of the extended chiral zero mode is opposite to the case in the absence of the Abrikosov vortices.
As a result, no fusion happens in this case of $N=3$ and odd $M_\text{u}+M_\text{l}$.

The above finding is generalized to other $N$'s. 
The even-odd behavior found in Sec.~\ref{Subsec:withoutVortices} occurs in the same way, when $M_\text{u}+M_\text{l}$ is even.
However, when $M_\text{u}+M_\text{l}$ is odd,  the behavior interchanges between even and odd $N$: In the case of even $N$, fusion between a localized Majorana zero mode inside the junction and the extended chiral Majorana zero mode appears in a range of $\varphi_0$, while fusion does not occur at any $\varphi_0$ in the case of odd $N$.
In Fig.~\ref{Fig:MFsVortices}(d) and (e), we provide the numerical result of the emergence of Majorana zero modes at $N=3$ and $N=2$ when $M_\text{u}+M_\text{l}=1$.

Next, we discuss the time evolution of the junction under the voltage bias $V_\textrm{DC}$ across the junction and the resulting AC Josephson effect in each case of $M_\textrm{u}$ and $M_\textrm{l}$.
In the case that both of $M_\textrm{u}$ and $M_\textrm{l}$ are even [see Fig.~\ref{Fig:MFsVortices}(a)], the $2n\pi$ fractional AC Josephson effect shows the same behaviour as in the case without Abrikosov vortices ($M_\textrm{u} = M_\textrm{l} = 0$), since this case has the same boundary condition of $\eta_a(x)$ in Eq.~\eqref{Eq:BC}; the Josephson current is $nT_\text{J}$-periodic with the period-elongation factor $n \ge 2$ determined by the DC voltage $V_\textrm{DC}$.

When both $M_\textrm{u}$ and $M_\textrm{l}$ are odd [see Fig.~\ref{Fig:MFsVortices}(b)], the boundary condition becomes $\eta_a(x=W) = \eta_a(x=-l)$.
Then, Majorana fermions evolves as $\gamma_{i=1,2,3}\mapsto\gamma_{i+1}$ and $\gamma_{4}\mapsto\gamma_1$   [cf. Eq.~\eqref{Eq:braiding}], since each of them passes the even number of branch cuts during time $T_\text{J} = h / (2e V_\textrm{DC})$. Meanwhile, the wave functions of Majorana zero modes $\psi_{a=\textrm{u,l}}$ hosted in the Abrikosov vortices in the upper ($a=\textrm{u}$) and lower ($a=\textrm{l}$) SCs evolve, during $T_\text{J}$, into $\psi_{\textrm{u}}(T_\text{J}) = -\psi_{\textrm{u}}(0)$ and $\psi_{\textrm{l}}(T_\text{J}) = \psi_{\textrm{l}}(0)$, respectively, in a gauge choice, where
the relative sign factor $-1$ originates from the SC phase evolution by $V_\textrm{DC}$.
In this case, the time-evolution operator $U$ has a different form from the case of  $M_\textrm{u} = M_\textrm{l} =0$  as 
\begin{widetext}
\begin{equation}
U
=
\frac{1}{\sqrt{2}}
\left(
\begin{array}{cccccccc}
 0 & 0 & e^{i \phi_-} & -e^{-i \phi_+} & 0 & 0 & 0 & 0 \\
 0 & 0 & e^{i \phi_+} & e^{-i \phi_-} & 0 & 0 & 0 & 0 \\
 i e^{-i \phi_-} & i e^{i \phi_+} & 0 & 0 & 0 & 0 & 0 & 0 \\
 i e^{-i \phi_+} & -i e^{i \phi_-} & 0 & 0 & 0 & 0 & 0 & 0 \\
 0 & 0 & 0 & 0 & 0 & 0 & e^{i \phi_-} & -e^{-i \phi_+} \\
 0 & 0 & 0 & 0 & 0 & 0 & e^{i \phi_+} & e^{-i \phi_-} \\
 0 & 0 & 0 & 0 & i e^{-i \phi_-} & i e^{i \phi_+} & 0 & 0 \\
 0 & 0 & 0 & 0 & i e^{-i \phi_+} & -i e^{i \phi_-} & 0 & 0 \\
\end{array}
\right) \label{S25}
\end{equation}
\end{widetext}
in basis $\{|011\rangle, |101\rangle, |000\rangle, |110\rangle, |010\rangle, |100\rangle, |001\rangle, |111\rangle \}$.  Here, $| n_{41} n_{32} n_{\text{vor}} \rangle \equiv (f_{41}^{\dagger})^{n_{41}}(f_{32}^{\dagger})^{n_{32}}|0_{41}0_{32}  n_{\text{vor}}  \rangle$, where  $n_{{\text{vor}}}=0,1$ denote the even and odd fermion parity defined by the vortices in the SCs, respectively (see the notation below Eq.~\eqref{Eq:Ue}). Although the time evolution operator conserves the total fermion parity $(-1)^{n_{41} + n_{32} + n_\textrm{vor}}$ during $T_\text{J}$,  the fermion parity of two complex fermions $f_{41}$ and $f_{32}$, which generate the Josephson current, changes in time.
It is because the fermion parity of the complex fermions formed by the vortices in the SCs changes during $T_\text{J}$; when both of $M_\textrm{u}$ and $M_\textrm{l}$ are odd, there always exists an odd number of the complex fermions in the SCs, one of which is formed by one vortex in the upper SC and another in the lower SC.
As a result, a modified $2n \pi$ fractional AC Josephson effect appears: The period-elongation factor of the Josephson current is an even integer $n \ge 2$; an odd integer period is not allowed.
This can be shown by using Eq.~\eqref{S25}.

When $M_\textrm{u} + M_\textrm{l}$ is odd [see Fig.~\ref{Fig:MFsVortices}(c)], there appears no Josephson current [see Fig.~\ref{Fig:MFsVortices}(d)] in the case of  $N=3$. It is enough to discuss the case that $M_\textrm{u}$ is odd and $M_\textrm{l}$ is even. In this case, there are four Majorana zero modes in one of the two configurations, (1) two localized Majorana zero modes inside the junction, one in the upper SC, and the other extending along the SC arcs, and (2) three localized Majorana zero modes inside the junction and the other in the upper SC.
The two configurations alternatively occur in time without any fusion between Majorana zero modes.
The period of the complex fermion states formed by the Majorana zero modes can be $nT_\text{J}$ with an integer $n \ge 2$.
However, the Josephson current does not occur since there is no fusion between the Majorana zero modes.

In the case of odd $M_\textrm{u} + M_\textrm{l}$, one can still see a fractional AC Josephson effect by reducing the magnetic field to have $N=2$ [see Fig.~\ref{Fig:MFsVortices}(e)]. Again, it is enough to discuss the case that $M_\textrm{u}$ is odd and $M_\textrm{l}$ is even. In this case, there are four Majorana fermions in one of the two configurations, (1') two localized Majorana zero modes inside the junction, one in the upper SC, and the other extending along the SC arcs, (2') one localized Majorana zero mode inside the junction, one in the upper SC, the other fused two Majorana fermions. The two configurations alternatively occur in time with fusions of Majorana zero modes, resulting in  interchanges of fusion pairs of the Majorana zero modes.
The time-evolution operator during one conventional Josephson period $T_\textrm{J}$ has the form of
\begin{equation}
U
=
\frac{1}{\sqrt{2}}
\left(
\begin{array}{cccc}
 e^{i\phi_+} & -e^{-i \phi_-} & 0 & 0 \\
 -i e^{i\phi_-} & -ie^{-i \phi_+} & 0 & 0 \\
 0 & 0 & -ie^{-\phi_+} & i e^{i \phi_-} \\
 0 & 0 & e^{-i \phi_-} & e^{i\phi_+} \\
\end{array}
\right)
\end{equation}
in the basis $\{|00\rangle, |11\rangle, |01\rangle, |10\rangle  \}$ of the complex fermions formed by the four Majoranas fermions. 
The Josephson current period can be $nT_\text{J}$ with the period-elongation factor which is an integer $n \ge 2$.

\section{Summary}\label{Sec:Summary}

We have studied the Josephson junction that can host localized Majorana zero modes inside the junction and the extended chiral Majorana zero modes along the boundary of the SC regions outside the junction. 
We have shown that the $2n \pi$ fractional AC Josephson effects occur in a realistic and experimentally feasible situations. Additionally, we have suggested an approach revealing the non-commutativity of the operations that braid the Majorana fermions of the junction.

%In this work, we focused on even-parity states. For odd parity...

Superposition of the basis states $|0_{41} 0 _{32} \rangle$ and $|1_{41} 1 _{32} \rangle$ formed by the four Majorana fermions constitutes a single-qubit state. The braiding operation of the four Majorana fermions can be considered as a rotation operation of the single qubit.
It will be interesting to extend our finding to multi-qubit states and multi-qubit operations by considering the case of $N \ge 7$ or in systems having multiple Josephson junctions formed by finite SCs.

%\bibitem{Clarke2} D. J. Clarke and Kirill Shtengel, Improved phase-gate reliability in systems with neutral Ising anyons, Phys. Rev. B {\bf 82}, 180519(R) (2010).

\section*{Acknowledgements}
This work was supported by Korea NRF (Grant No. 2016R1A5A1008184; H.-S.S. and S.-J. C.) and IBS-R024-D1 (S.-J.C).

\section*{Appendix A: Expressions of $\Phi_a(X_a,Y_a)$}\label{AppendixA}
The wave function $\Phi_{a}(X_a,Y_{a})$ of the chiral mode $\eta_a$ along the arcs of the SC regions (see Sec.~\ref{Subsec:Hamiltonian}) is written, up to normalization, as
\begin{widetext}
\begin{eqnarray}
\Phi_{a}(X_a,Y_{a}) &=& e^{-\frac{Y_{a}^2}{2 l_B^2}}
\left(
\begin{array}{c}
s_{a} \tilde{\mu} H_{\tilde{\mu}^2/2-1}\left(- \frac{s_{a} Y_{a}}{l_B}\right) e^{-i\theta/2} \\
 H_{\tilde{\mu}^2/2}\left(- \frac{s_{a}Y_{a}}{l_B}\right) e^{i\theta/2} \\
 H_{\tilde{\mu}^2/2}\left(- \frac{s_{a}Y_{a}}{l_B}\right) e^{-i\theta/2} \\
-s_{a} \tilde{\mu} H_{\tilde{\mu}^2/2-1}\left(- \frac{s_{a}Y_{a}}{l_B}\right) e^{i\theta/2}
\end{array}
\right), \,\,\,\, s_{a}Y_{a} < 0 \\
\Phi_{a}(X_a,Y_a) &=& e^{-\frac{1}{\hbar v_F}\int_{0}^{|Y_{a}|}\Delta_0(Y_{a}')dY_{a}'}
\left[
A_{a} \left(
\begin{array}{c}
\cos \left(\frac{\mu Y_{a}}{\hbar v_F}\right) e^{-i\theta/2} \\
-\sin \left(\frac{\mu Y_{a}}{\hbar v_F}\right) e^{i\theta/2} \\
-\sin \left(\frac{\mu Y_{a}}{\hbar v_F}\right) e^{-i\theta/2} \\
-\cos \left(\frac{\mu Y_{a}}{\hbar v_F}\right) e^{i\theta/2} \\
\end{array}
\right)
+
B_{a}\left(
\begin{array}{c}
\sin \left(\frac{\mu Y_{a}}{\hbar v_F}\right) e^{-i\theta/2} \\
\cos \left(\frac{\mu Y_{a}}{\hbar v_F}\right) e^{i\theta/2} \\
\cos \left(\frac{\mu Y_{a}}{\hbar v_F}\right) e^{-i\theta/2} \\
-\sin \left(\frac{\mu Y_{a}}{\hbar v_F}\right) e^{i\theta/2} \\
\end{array}
\right)
\right], \,\,\,\, s_{a}Y_{a} > 0, \label{S4}
\end{eqnarray}
\end{widetext}
where $s_{\textrm{u}}=1$, $s_{\textrm{l}}=-1$, $\tilde{\mu} \equiv \mu l_B/(\hbar v_F)$, $l_B=\sqrt{\hbar/(eB)}$ is the magnetic length, and $\theta$ is an angle defined in Fig.~\ref{Fig:setup}. The coefficients $A_{a} = s_{a} \tilde{\mu} H_{\tilde{\mu}^2/2-1}(0)$ and $B_{a} = H_{\tilde{\mu}^2/2}(0)$ are determined by matching wave functions at $Y_a = 0$ along the arcs of the SC regions. 
Inside the junction, the wave functions of the chiral modes can be written as
$ \eta_a (x) = \int \Psi^{\dagger}(x,y)\Phi_a (x,y)dy$ at position $x$,
\begin{equation}
\Phi_{a}(x,y) = 
\frac{e^{-|y|/\xi}}{2\sqrt{\xi}}
\left(
\begin{array}{c}
s_{a} \cos \frac{\mu y}{\hbar v_F} + \sin \frac{\mu y}{\hbar v_F} \\
-s_{a} \sin \frac{\mu y}{\hbar v_F} + \cos \frac{\mu y}{\hbar v_F} \\
-s_{a} \sin \frac{\mu y}{\hbar v_F} + \cos \frac{\mu y}{\hbar v_F} \\
-s_{a} \cos \frac{\mu y}{\hbar v_F} - \sin \frac{\mu y}{\hbar v_F} \\
\end{array}
\right). \label{S4}
\end{equation} 
%where $\xi\equiv\hbar v_F/\Delta_0$.
%
%\section*{Appendix B: Discretization of $H$}\label{AppendixB}
%(Lattice Hamiltonian of $H_\text{DBdG}$)
%$t_x = it\sigma_x - \frac{2t}{\sqrt{3}}\sigma_z$, $t_y = it\sigma_y - \frac{2t}{\sqrt{3}}\sigma_z$, and on-site potential of $\frac{4t}{\sqrt{3}}\sigma_z$ accomplish the single Dirac cone upto the third order of $k_x$ and $k_y$.
%
%We introduce the lattice Hamiltonian $H_{\text{lattice}}(t)$ of $H(t)$, which is $H_{\text{lattice}}(t)=\sum_{n=-N_\text{arc}}^{N_\text{J}}\left(\Gamma^\top_n h_n(t) \Gamma_n + \Gamma^\top_{n+1} h_x \Gamma_n + h.c \right)$, where $\Gamma_n = \left( \eta_\text{u}(x=nc) ,\,\, \eta_\text{l}(x=nc) \right)^\top$, $h_n(t) = -\frac{\hbar v_\text{J}}{c}\sigma_y + m(x=nc,t)\sigma_y, h_x = \frac{\hbar v_\text{J}}{2c}\left(\sigma_y + i\sigma_z \right)$, and $\sigma_{y,z}$ are the Pauli matrices in the basis of the chiral Majorana modes. We can use $v_\text{J}$ for the arcs of SCs, since $H_\text{arc}=-iv_\text{arc}\sum_{a}\int_{-l_a}^0\eta_a(X_a)\partial_{X_a}\eta_a(X_a)dX_a=-iv_\text{J}\sum_{a}\int_{-(v_\text{J}/v_\text{arc})l_a}^0\tilde{\eta_a}(X_a)\partial_{X_a}\tilde{\eta_a}(X_a)dX_a$ for the rescaling of $X_a\mapsto X_a v_\text{arc}/v_\text{J}$. $N_\text{J}=W/c$ ($N_\text{arc}=(v_\text{J}/v_\text{Arc})L/c$) is the number of lattice sites along the junction (arc), and $c$ is the lattice constant. $\Gamma_{N_\text{J}+1} = -\Gamma_{-N_\text{arc}}$, whose minus sign comes from the anti-periodic boundary condition of the continuum Hamiltonian.

\section*{Appendix B: Derivation of $I_\textrm{MF}$}\label{AppendixCurrent}

We derive the relation of $I_\textrm{MF} = (1/ V_\textrm{DC}) dE_\textrm{MF}/ dt$ used in Sec.~\ref{Subsec:FJ}. The contribution  $I_\textrm{MF}$ of the four Majorana fermions to the Josephson current is obtained as  $I_\text{MF}\equiv\int_0^W dx \langle \hat{J}(x)\rangle$, where $\hat{J}(x)\equiv s\frac{e\Delta_0}{\hbar}\Gamma^\top(x)\sigma_y\Gamma(x)$ is the Josephson-current density operator in the basis space of the chiral Majorana modes $(\eta_\textrm{u},\eta_\textrm{l})$, $s= \text{sgn}[\sin(N\pi x/W-eV_\text{DC}t/\hbar)]$, and the expectation value $\langle \cdots \rangle$ is obtained with the states formed by the Majorana fermions. The matrix element of $\hat{J}(x)$ is derived as
\begin{equation*}
ev_F
\left(
\begin{array}{cc}
\zeta_\text{u}^{\dagger}(x) s_y \zeta_\text{u}(x) & \zeta_\text{u}^{\dagger}(x) s_y \zeta_\text{l}(x) \\
\zeta_\text{l}^{\dagger}(x) s_y \zeta_\text{u}(x) & \zeta_\text{l}^{\dagger}(x) s_y \zeta_\text{l}(x)
\end{array}
\right),
\end{equation*}
%by projecting the corresponding operator $J_y = e v_F s_y$ written in the Nambu basis $\left( \psi_{\uparrow}(\textbf{r}), \psi_{\downarrow}(\textbf{r}), \psi_{\downarrow}^{\dagger}(\textbf{r}), - \psi_{\uparrow}^{\dagger}(\textbf{r}) \right)^\top$ into chiral Majorana wave functions 
where $\zeta_\text{u,l}(x)\equiv\Phi_\text{u,l}(x,y=0)$ and $s_y$ is the $y$-component of the Pauli matrices for electron spins.
% around the position $x$ satisfying $\Delta \varphi(x,t)= \pi$ along the line of $y=0$, where $\zeta_\text{u}^{\dagger}(x) = \frac{1}{2}\sqrt{\frac{\Delta_0}{\hbar v_F}}(1,\,1,\,s,\,-s)$ and $\zeta_\text{l}^{\dagger}(x) = \frac{1}{2}\sqrt{\frac{\Delta_0}{\hbar v_F}}(-s,\,s,\,1,\,1)$ as in Eq.~\eqref{S4}.
%Similarly we also obtain $\hat{J}(x)$ around the positions $x$ of $\Delta\varphi(x,t) = 3\pi, 5\pi,\cdots$. Then for $x \in [0,W]$, $\hat{J}(x)\equiv s\frac{e\Delta_0}{\hbar}\Gamma^\top(x)\sigma_y\Gamma(x)$ is found in the basis space of the chiral Majorana modes $(\eta_\textrm{u},\eta_\textrm{l})$.

Next, $\frac{1}{V_\text{DC}}\frac{dE_\text{MF}}{dt}$ is evaluated with the state $|Q(t) \rangle$ of the two complex fermions formed by the four Majorana fermions $\gamma_{k=1,2,3,4}$,
\begin{widetext}
\begin{eqnarray*}
\frac{1}{V_\text{DC}}\frac{dE_\text{MF}}{dt}
&=&\frac{1}{V_\text{DC}} \langle Q(t)|\frac{dH(t)}{dt} |Q(t) \rangle \\
&=&\frac{e\Delta_0}{\hbar}\int_0^W dx \langle Q(t)| \Gamma^\top(x) \sin\left(\frac{N\pi x}{W}-\frac{eV_\text{DC}t}{\hbar}\right)\sigma_y \Gamma(x)|Q(t) \rangle  \\
&\approx&\int_0^W dx \langle Q(t)|s\frac{e\Delta_0}{\hbar}\Gamma^\top(x)\sigma_y\Gamma(x) |Q(t) \rangle \\
&=& \int_0^W dx\langle\hat{J}(x)\rangle.
\end{eqnarray*}
\end{widetext}
We use the Feynman-Hellman theorem in the first line of the derivation. In the approximation of the third line, we use $\sin\left(\frac{N\pi x_k}{W}-\frac{eV_\text{DC}t}{\hbar}\right) = s$ that is applicable when $\lambda\ll\frac{W}{N}$, where $s$ is the sign factor defined above.

\end{document}